\documentclass[12pt]{iopart}
\expandafter\let\csname equation*\endcsname\relax
\expandafter\let\csname endequation*\endcsname\relax
\usepackage{amsmath}
\usepackage{cite}
\usepackage{graphicx}
\usepackage{bm}
\usepackage{hyperref}
\hypersetup{colorlinks=true, citecolor=blue, urlcolor=blue, linkcolor=blue}
\usepackage{subfigure}
\begin{document}

\title[]{Stationary and moving bright solitons in Bose-Einstein condensates with spin-orbit coupling in a Zeeman field}

\author{Juntao~He\textsuperscript{1} and Ji~Lin\textsuperscript{1,*}}
\address{\textsuperscript{1} Department of Physics, Zhejiang Normal University, Jinhua 321004, China
\\ \textsuperscript{*} Authors to whom any correspondence should be addressed.}
\ead{linji@zjnu.edu.cn}
\vspace{10pt}

\begin{abstract}
We explore the existence and stability domains of stationary and moving bright solitons in spin-orbit-coupled spin-1 Bose-Einstein condensates in an external Zeeman field. Two families of bell-shape bright solitons (plane-wave phases) and two families of stripe-shape bright solitons (standing-wave phases) are obtained. We find a relation between the existence of these bright solitons and the single-particle energy spectrum, as well as the requirements of their existence for atomic interactions. The stability of four families of bright solitons is systematically analyzed using the linear stability analysis method. For two families of plane-wave bright solitons, they are unstable only when the strength of the Zeeman field is above a critical value. The critical value is affected by the spin-orbit coupling and hardly affected by atomic interactions. When the velocities of solitons are zero (nonzero), the critical values of these two families are equal (unequal). For two families of standing-wave bright solitons, their stability domains are identical in the absence of the Zeeman field, and they are unstable when the ferromagnetic interaction is strong. In the presence of the Zeeman field, their stability domains are complementary and complex, and one family of standing-wave bright solitons can exist stably in the area with stronger ferromagnetic interaction. Furthermore, we find the collisions of stable bright solitons with different velocities can generate intriguing dynamics.
\end{abstract}

\vspace{2pc}

\submitto{\NJP}
%
%
%

\section{Introduction}
The spin-orbit coupling (SOC) has been realized in spinor Bose-Einstein condensates (BECs) by dressing atomic spin states with lasers \cite{lin_spinorbit-coupled_2011,luo_tunable_2016}, which is equivalent to an equal weight mixing of Rashba \cite{bychkov_oscillatory_1984} and Dresselhaus \cite{dresselhaus_spin-orbit_1955} SOC. Due to its tunability, a spinor BEC with SOC is a good platform for studying condensed matter effects, such as topological insulators \cite{qi_topological_2011} and spin Hall effects \cite{sinova_spin_2015}. In such a system, there are various intriguing ground state phases, such as the zero-momentum (ZM), plane-wave (PW), standing-wave (SW), triangular-lattice, and square-lattice phases \cite{wang_spin-orbit_2010,wen_ground_2012,ho_bose-einstein_2011,li_quantum_2012,lan_raman-dressed_2014,natu_striped_2015,sun_interacting_2016,campbell_magnetic_2016,maeland_plane-_2020,adhikari_multiring_2021,adhikari_symbiotic_2021}. Among them, the PW and SW phases exhibit nonzero magnetizations \cite{chen_elementary_2022} and supersolid properties \cite{li_stripe_2017,sanchez-baena_supersolid_2020}, respectively. They are unique and exotic in quasi-one-dimensional systems with SOC. In general, these phases behave as bright solitons when atomic interactions are attractive \cite{wan_solitons_2019}. Studying bright solitons is essential for the studies of quantum states and dynamic phenomena \cite{wen_motion_2016,kartashov_bloch_2016,kartashov_solitons_2019,qiu_dynamics_2023}.

Bell-shape PW bright solitons and stripe-shape SW bright solitons spontaneously break the symmetries of time inversion and spatial continuity, respectively. They have been extensively studied in spin-1/2 BECs with SOC when considering the Zeeman effect \cite{xu_bright_2015,xu_bright_2013,achilleos_matter-wave_2013,kartashov_bose-einstein_2014,sakaguchi_vortex_2016,li_moving_2016,kartashov_multidimensional_2020,kartashov_solitons_2017,li_solitary_2021}, which affects the profiles of solitons and enriches the magnetic behaviors of the system \cite{zhao_topological_2015,zapf_bose-einstein_2014}. Due to the combined action of SOC and the Zeeman effect, moving PW and SW bright solitons exhibit exotic inelastic collision phenomena \cite{kartashov_solitons_2017,li_solitary_2021}. Meanwhile, the absence of Galilean invariance makes it challenging to construct moving bright solitons whose profiles are constant during the evolution. \cite{sun_bright_2020,tononi_quantum_2019}. Compared with spin-1/2 BECs, the spin-1 BEC systems have more adjustable parameters and contain more physical properties \cite{kawaguchi_spinor_2012}. The relevant research about spin-1 BECs with SOC mainly focuses on the situation without the Zeeman effect \cite{gautam_mobile_2015,adhikari_phase_2019,gautam_vortex-bright_2017,he_multi-type_2022,qi_soliton_2023}. When the Zeeman effect is present, stationary and moving bright solitons in spin-1 BECs with SOC have not been systematically explored \cite{li_stationary_2018}, especially SW bright solitons. As we know, the mean-field ground state of a spin-1 BEC with SOC favors a PW (SW) bright soliton for the ferromagnetic (antiferromagnetic) interaction \cite{wang_spin-orbit_2010,wen_ground_2012}. This extends to several questions of whether it is possible to find PW (SW) bright solitons in the antiferromagnetic (ferromagnetic) area, what the specific existence domains of PW and SW bright solitons are, and how the Zeeman field affects the existence domains of bright solitons. Due to the noise in experiments, the stability of bright solitons determines whether they can be observed and studied in experiments. Therefore, discussing the stability domains of bright solitons is necessary.

In this work, we systematically investigate stationary and moving bright solitons in spin-1 BECs with SOC in a Zeeman field, focusing mainly on the bright solitons corresponding to the PW phase and the SW phase. Without the Zeeman field, we establish a connection between the single-particle energy spectrum and exact soliton solutions, whereby we obtain the existence domains of stationary PW and SW bright solitons. Similarly, the existence domains of moving PW and SW bright solitons are also obtained, which are stationary solitons in the moving frame. Besides, we analyze the stability of these bright solitons by linear stability analysis. Furthermore, we investigate the existence and stability domains of PW and SW bright solitons when the Zeeman field exists, based on the single-particle energy spectrum and a large number of numerical solutions. We find that the linear Zeeman effect is advantageous (disadvantageous) to the stability of SW (PW) bright solitons. Stable SW and PW bright solitons in both ferromagnetic and antiferromagnetic BECs are found. Finally, we discuss some interesting collision dynamics of stable bright solitons.

This paper is organized as follows. We first give the model in section~\ref{sec2} and calculate the single-particle energy spectrum in section~\ref{sec3}. In the absence or presence of a Zeeman field, we investigate the existence and stability domains of stationary and moving bright solitons analytically and numerically in section~\ref{sec4}, and discuss some collision dynamics of two stable bright solitons in section~\ref{sec5}. Finally, we conclude our results in section~\ref{sec6}.

\section{Model}\label{sec2}
First of all, we assume that the longitudinal trapping frequency $\omega_x$ is much smaller than the transverse trapping frequency $\omega_\perp$. In this case, the wave function of transverse is the ground state of the harmonic oscillator. Therefore, a BEC can be regarded as a quasi-one-dimensional system. Considering the equal weight mixing of Rashba and Dresselhaus SOC, a spin-1 BEC in a Zeeman field can be described by quasi-one-dimensional three-component Gross-Pitaevskii equations with SOC \cite{kawaguchi_spinor_2012}. After dimensionless \cite{gautam_mobile_2015,adhikari_phase_2019}, the equations become
\begin{eqnarray}
\begin{aligned}\label{eq1}
\mathrm{i}\frac{\partial \psi_{\pm1}}{\partial{t}}=&\left(-\frac{1}{2} \frac{\partial^{2}}{\partial{x}^{2}}+c_0n+c_2\left(n_{\pm1}+n_0-n_{\mp1}\right)\right)\psi_{\pm1}\\&+c_{2} \psi_{0}^{2} \psi_{\mp1}^{*}-\frac{\mathrm{i}\gamma}{\sqrt{2}}\frac{\partial\psi_0}{\partial x}+(q\mp p)\psi_{\pm1},\\
\mathrm{i}\frac{\partial \psi_{0}}{\partial{t}}=&\left(-\frac{1}{2} \frac{\partial^{2}}{\partial{x}^{2}}+c_0n+c_2\left(n_{+1}+n_{-1}\right)\right)\psi_{0}\\&+2c_{2} \psi_{0}^{*} \psi_{+1}\psi_{-1}-\frac{\mathrm{i}\gamma}{\sqrt{2}}\left(\frac{\partial\psi_{+1}}{\partial x}+\frac{\partial\psi_{-1}}{\partial x}\right),
\end{aligned}
\end{eqnarray}
where $\gamma$ is the strength of SOC (controlled by two lasers) and $\psi_{j}$ (with $j=+1,0,-1$) are the wave functions of the three hyper-fine spin components of a spinor BEC. The $n_{j}=|\psi_j|^2$ are the component densities, while $n=\sum_j|\psi_j|^2$ is the total particle density. In addition, $c_0$ and $c_2$ are the effective constants of mean-field and spin-exchange interactions, respectively. They are connected with two-body s-wave scattering lengths $a_0$ and $a_2$ for total spin 0, 2: $c_0\propto\left(a_0+2a_2 \right)/3$ and $c_2\propto\left(a_2-a_0\right)/3$, which can be adjusted experimentally by the Feshbach resonance technology \cite{papoular_microwave-induced_2010}. This allows us to adjust whether the system is ferromagnetic ($c_2<0$) or antiferromagnetic ($c_2>0$) and whether the atomic interaction is attractive or repulsive, which are critical for the types of solitons allowed in the system. Moreover, $p=-\mu_{\mathrm{B}} gB$ and $q=g^2\mu^2_{\mathrm{B}}B^2/\Delta E_{\mathrm{hf}}$ represent the strength of the linear and quadratic Zeeman effects, respectively. Here $\mu_{\mathrm{B}}$ is the Bohr magneton, $B$ is the strength of the magnetic field, $g$ is the Land\'{e} g-factor, and $\Delta E_{\mathrm{hf}}$ is the hyperfine energy splitting between the initial and intermediate energies. Therefore, both $p$ and $q$ can be positive or negative. In fact, upon a pseudo-spin rotation, our model is suitable for a spin-1 BEC system with Raman-induced SOC \cite{lan_raman-dressed_2014}.

\section{Single-particle energy spectrum}\label{sec3}
When we discuss the multi-particle system, it is necessary to analyze the single-particle energy spectrum, which is the relationship between the momentum and energy of a single particle \cite{ravisankar_effect_2021}. It can be obtained by solving the eigenvalues of single-particle Hamiltonian or the PW solutions of linear Schr\"{o}dinger equations. In a BEC, almost all particles are in the same state, while the state corresponds to a point in the energy spectrum when the interactions are ignored. Considering the attraction (repulsion) interaction, the energy of the system decreases (increases). In other words, if the energy of the system is below (above) the energy spectrum, the wave function of the system is a bright (dark) soliton. Besides, the derivative of the energy spectrum determines the group velocity of the system, which is the velocity of the corresponding soliton. Therefore, the single-particle energy spectrum plays an important role in our study of matter-wave solitons in spinor BECs with SOC.

\begin{figure}[h]
	\flushright
	\subfigure{\label{fig1a}
		\begin{minipage}[t]{0.35\linewidth}
			\includegraphics[width=5cm]{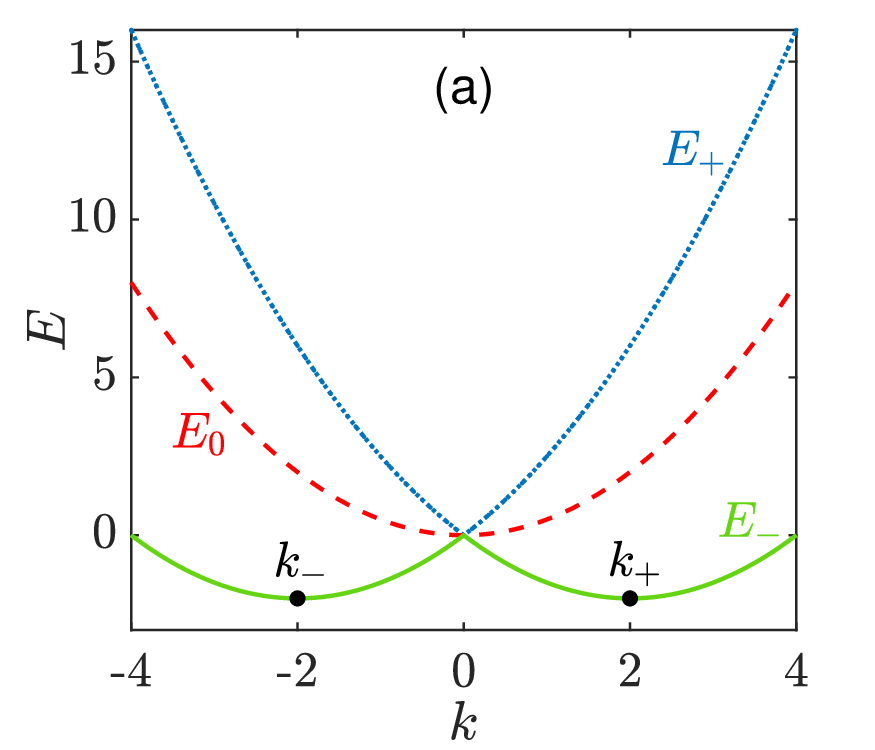}
	\end{minipage}}
	\subfigure{\label{fig1b}
		\begin{minipage}[t]{0.35\linewidth}
			\includegraphics[width=5cm]{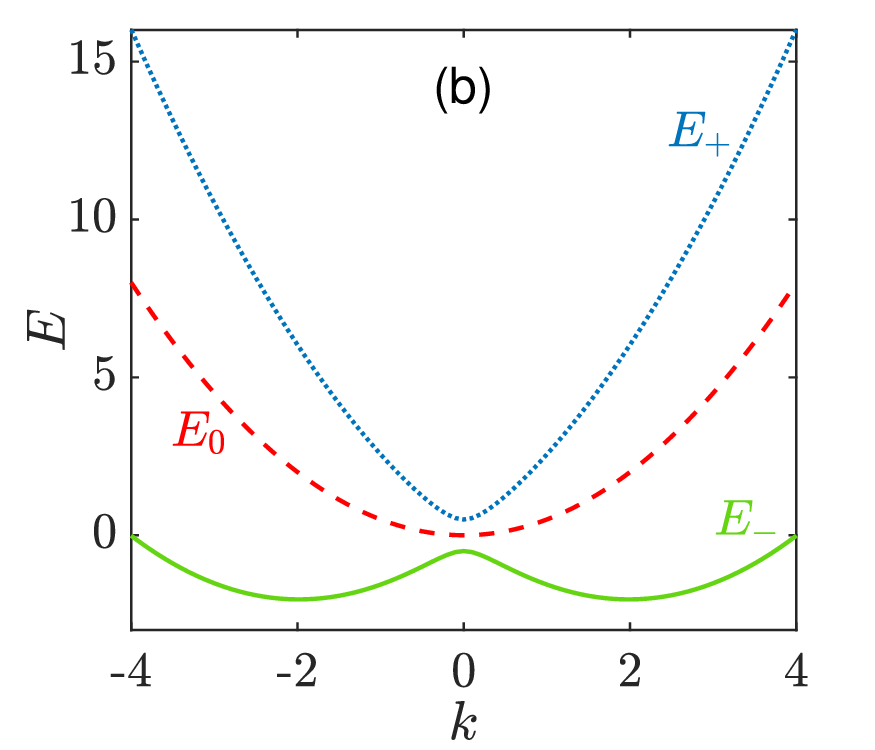}
	\end{minipage}}
	\caption{Three branches of the single-particle energy spectrum. (a) $p=0$. (b) $p=0.5$. Here $\gamma=2$ and $q=0$. $k_{\pm}$ are the momentum values corresponding to the two local minimums of $E_-$.}
\end{figure}

The expression of the energy spectrum is very complex when both $p$ and $q$ are considered, so we discuss their influence on the energy spectrum separately. By solving the eigenvalues of the single-particle Hamiltonian of equation~\eqref{eq1}, we obtain the energy spectrum when $q=0$ or $p=0$:
\begin{subequations}
	\begin{eqnarray}
	&p:&~~E_0=\frac{k^2}{2},~~E_{\pm}=\frac{k^2}{2}\pm\sqrt{\gamma^2 k^2+p^2},\label{eq3a} \\
	&q:&~~E_0=\frac{k^2}{2}+q,~~E_{\pm}=\frac{k^2}{2}+\frac{q}{2}\pm\sqrt{\gamma^2k^2+\frac{q^2}{4}}\label{eq3b},
	\end{eqnarray}
\end{subequations}
where $k$ is the momentum, and $E_{\pm,0}$ are three energy branches in momentum space. When $q=0$ and $p=0$, we find that $\gamma$ makes the lowest branch of the energy spectrum have two different local minimums, as shown in figure~\ref{fig1a}.
Their momentum values are $k_+$ and $k_-$, respectively. The two local minimums mean that the system has multiple stationary states. When all atoms in the BEC occupy a single momentum ($k_+$ or $k_-$), the system behaves as the PW phase ($\sqrt{\rho(x)}\mathrm{exp}\left(\mathrm{i}k_{\pm}x\right)$), where $\rho(x)$ is a real function. In particular, it behaves as the ZM phase ($\sqrt{\rho(x)}$) when the single momentum is zero. Besides, when atoms occupy two momentums ($k_+$ and $k_-$), the system behaves as the SW phase (the combination of two PW phases).

Comparing equations~\eqref{eq3a} and \eqref{eq3b}, $p$ and $q$ have similar effects on the structure of the energy spectrum, and $p$ makes the three branches separate from each other, as shown in figure~\ref{fig1b}. As long as the strength of $p$ and $q$ is relatively not too strong, they only shift the phase boundary between these phases, and not affect the essence of the phases \cite{wang_spin-orbit_2010}. We only consider the case of $q=0$ because $q$ is generally much smaller than $p$. 

\section{Matter-wave solitons and their existence and stability domains}\label{sec4}
\subsection{Without a Zeeman field}
Matter-wave solitons exist in every branch of the single-particle energy spectrum. However, only ZM solitons exist in the upper branches, and their stability is relatively weak. We will mainly consider the solitons corresponding to the lowest branch $E_-$. At first, we discuss the solitons in a spin-1 BEC with SOC in the absence of Zeeman field $(p=0)$. In this case, we can simplify equation~\eqref{eq1} into the Nonlinear Schr\"{o}dinger equation by some transformations \cite{he_multi-type_2022}, which is an integrable equation and has various exact solutions \cite{kanna_exact_2003}. The Nonlinear Schr\"{o}dinger equation is
\begin{eqnarray}\label{eq4}
\mathrm{i}\frac{\partial u}{\partial t}+\frac{1}{2}\frac{\partial^2 u}{\partial x^2}+\sigma |u|^2u=0,
\end{eqnarray}
where $u$ is the complex function of $x$ and $t$, and $\sigma=\pm1$ represent attractive and repulsive interactions, respectively. The key transformations between equations~\eqref{eq1} and \eqref{eq4} are
\begin{subequations}
	\begin{eqnarray}
	&\left(\begin{array}{c}\label{eq5a}
	\psi_{+1} \\ \psi_{0} \\ \psi_{-1} \end{array}\right)&=\frac{\mathrm{e}^{\frac{\mathrm{i}}{2}\gamma^2 t \pm \mathrm{i}\gamma x}}{2\sqrt{g_{\mathrm{p}}/\sigma}}u
	\left(\begin{gathered}
	1\\ \mp\sqrt{2} \\ 1 \end{gathered}\right),
	\\
	&\left(\begin{array}{c}\label{eq5b}
	\psi_{+1} \\ \psi_{0} \\ \psi_{-1} \end{array}\right)&=\frac{\mathrm{e}^{\frac{\mathrm{i}}{2}\gamma^2 t}}{2\sqrt{g_{\mathrm{s}}/\sigma}}u
	\left(\begin{gathered}
	\mathrm{e}^{\mathrm{i}\gamma x}\pm \mathrm{e}^{-\mathrm{i}\gamma x}\\ -\sqrt{2}\left(\mathrm{e}^{\mathrm{i}\gamma x}\mp \mathrm{e}^{-\mathrm{i}\gamma x}\right) \\ \mathrm{e}^{\mathrm{i}\gamma x}\pm \mathrm{e}^{-\mathrm{i}\gamma x} \end{gathered}\right).
	\end{eqnarray}
\end{subequations}
Here, $g_{\mathrm{p}}=-\left(c_0+c_2\right)$ and $g_{\mathrm{s}}=-2c_0$. We notice that there are two PWs $\left(\mathrm{exp}(\pm \mathrm{i}\gamma x)\right)$ in these transformations. In addition, the momentums $k_{\pm}$ corresponding to the local minimums of lowest branch are equal to $\pm\gamma$ in the absence of Zeeman field. It means equations~\eqref{eq5a} and \eqref{eq5b} represent the wave functions of the atoms occupying a single momentum and two momentums, respectively. Therefore, they correspond to the exact solutions of the PW phase and the SW phase of equation~\eqref{eq1}. 

\begin{figure}[h]
	\flushright
	\subfigure{\label{fig2a}
		\begin{minipage}[t]{0.35\linewidth}
			\includegraphics[width=5cm]{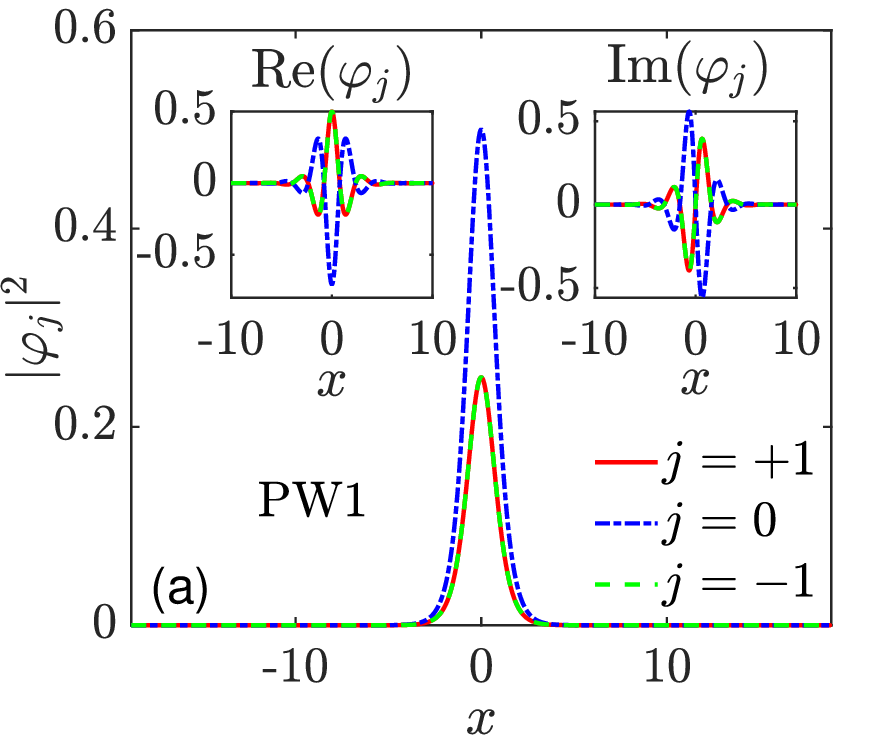}
	\end{minipage}}
	\subfigure{\label{fig2b}
		\begin{minipage}[t]{0.35\linewidth}
			\includegraphics[width=5cm]{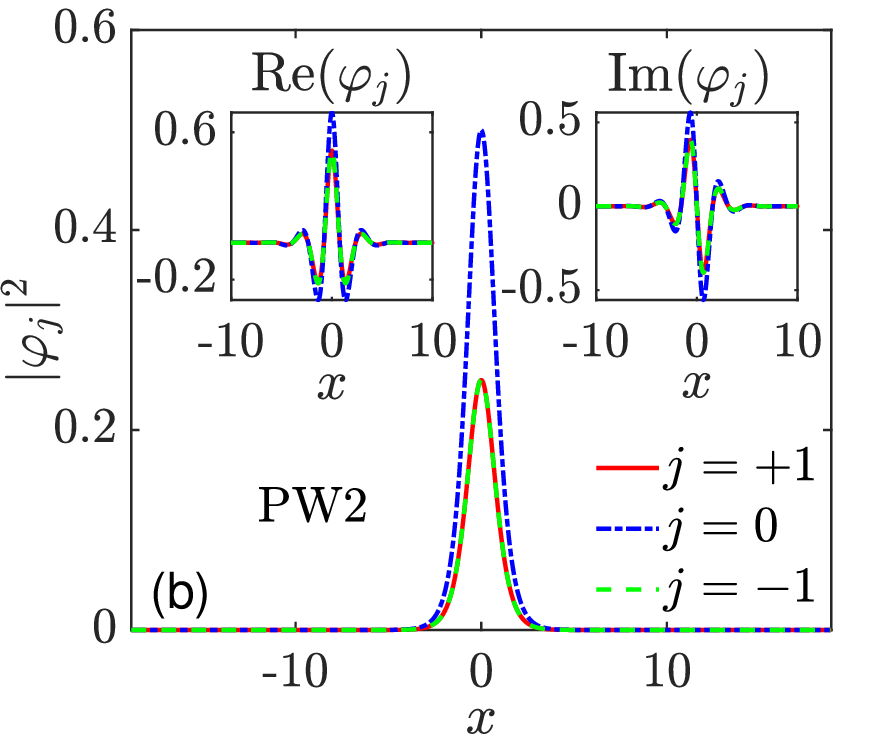}
	\end{minipage}}
	\subfigure{\label{fig2c}
		\begin{minipage}[t]{0.35\linewidth}
			\includegraphics[width=5cm]{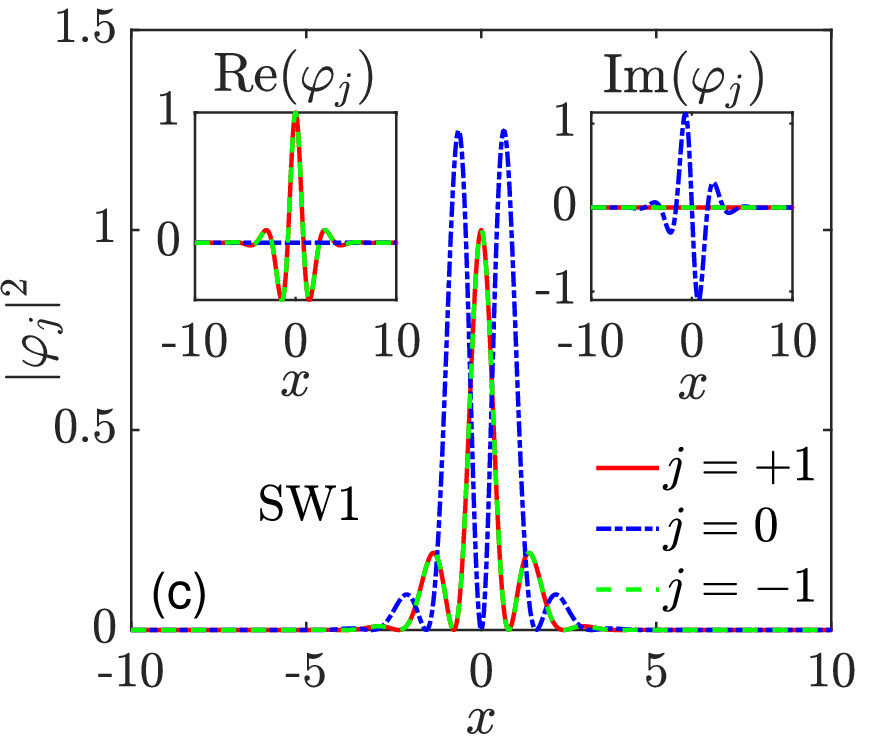}
	\end{minipage}}
	\subfigure{\label{fig2d}
		\begin{minipage}[t]{0.35\linewidth}
			\includegraphics[width=5cm]{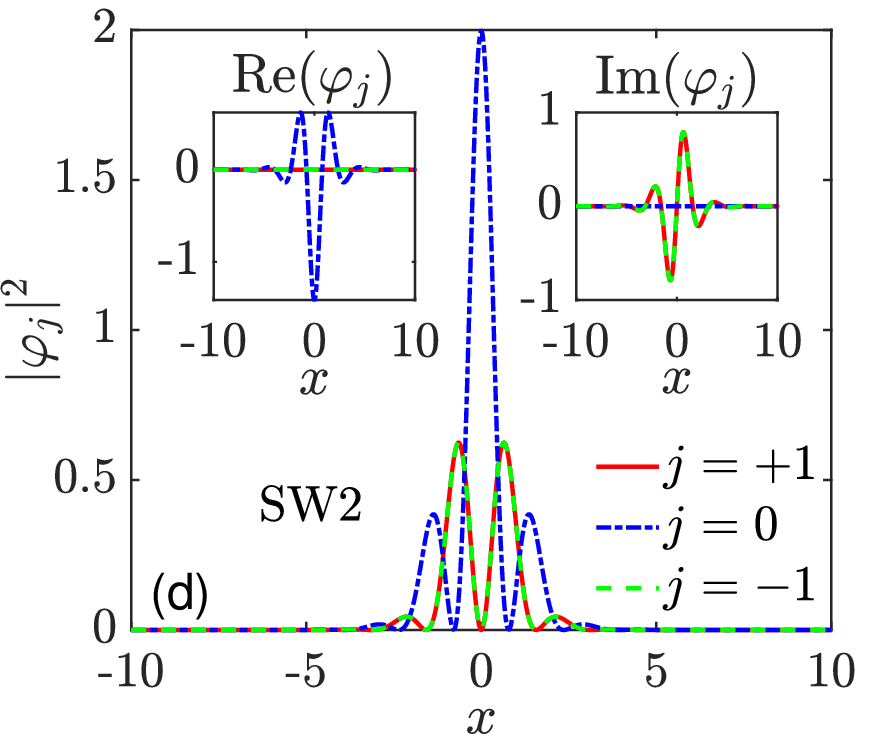}
	\end{minipage}}
	\caption{The four types of stationary bright solitons when $p=0$. The plane wave solitons (PW1, PW2) for $c_0=0.5$, $c_2=-1.5$ are in (a) and (b). The standing wave solitons (SW1, SW2) for $c_0=-0.5$, $c_2=1.5$ are in (c) and (d). The left and right subgraphs represent the real and imaginary parts of the solitons, respectively. Here, $k_1=1$ and $\gamma=2$.}\label{fig2}
\end{figure}

With the help of equations~\eqref{eq5a} and \eqref{eq5b}, we can obtain a lot of exact analytical solutions when $p=0$. We are mainly concerned about stationary-state solutions, especially bright solitons. Thus we consider the attractive interaction ($\sigma=1$), and a bright soliton solution of equation~\eqref{eq4} is
\begin{eqnarray}\label{eq6}
u=k_1{\rm sech}\left(k_1\left( x-vt\right) \right)\mathrm{e}^{\mathrm{i}vx+\mathrm{i}\left(k_1^2-v^2\right)t/2},
\end{eqnarray}
where $k_1$ and $v$ are real parameters, representing the amplitude and velocity of the bright soliton, respectively. Substituting equation~\eqref{eq6} into equations~\eqref{eq5a} and \eqref{eq5b}, we derive four types of bright solitons which are stationary states when $v=0$. For convenience, the solitons mentioned later are defaulted to bright solitons. Drawing their profiles at $t=0$ in figure~\ref{fig2}, we can observe their structures and differences. In figures~\ref{fig2a} and \ref{fig2b}, the PW solitons (PW1 and PW2, respectively) have bell-shaped structures, and their component densities $n_j$ are the same. In addition, the SW solitons in figures.~\ref{fig2c} and \ref{fig2d} (SW1 and SW2, respectively) have stripe-shaped structures.
The four types of solitons are unique and concerned in systems with SOC. As we know, a stationary-state solution can be divided into space function $\varphi(x)$ and time function $\mathrm{exp}\left(-\mathrm{i}\mu t\right)$. Here, $\mu$ is the chemical potential, and we derive the chemical potentials of the four types of solitons $\mu=-\left(k_1^2+\gamma^2 \right)/2$. Additionally, the local minimums of $E_-$ are all $-\gamma^2/2$. We conclude that the widths and amplitudes $k_1$ of bright solitons are determined by the difference between the local minimum of the energy spectrum and the chemical potential $\mu$. In fact, this is a universal law that holds true in other systems.

\begin{figure}[h]
	\flushright
	\subfigure{\label{fig3a}
		\begin{minipage}[t]{0.35\linewidth}
			\includegraphics[width=5cm]{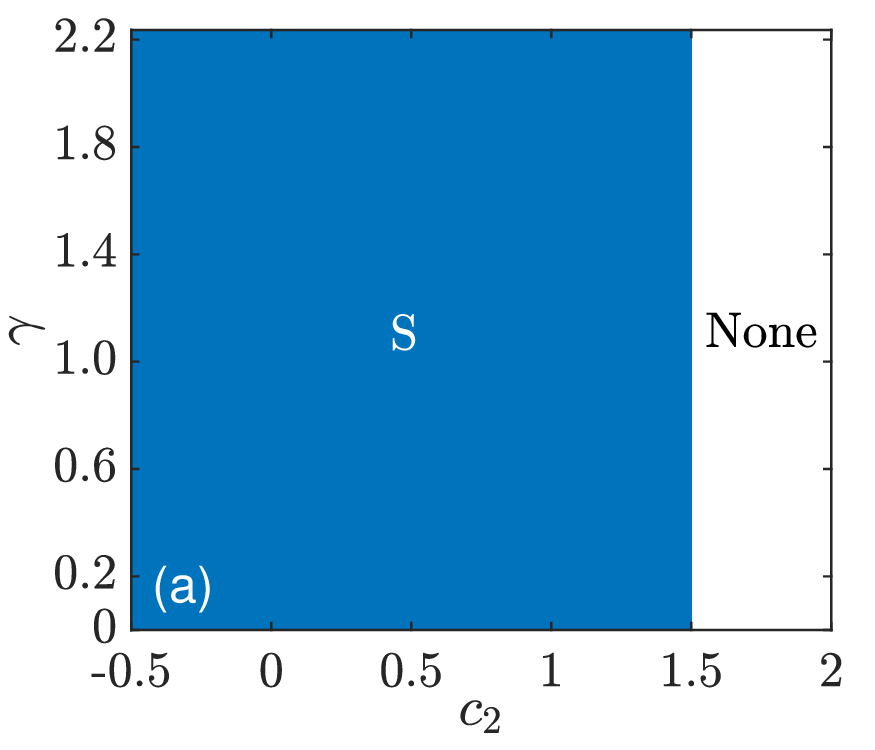}
	\end{minipage}}
	\subfigure{\label{fig3b}
		\begin{minipage}[t]{0.35\linewidth}
			\includegraphics[width=5cm]{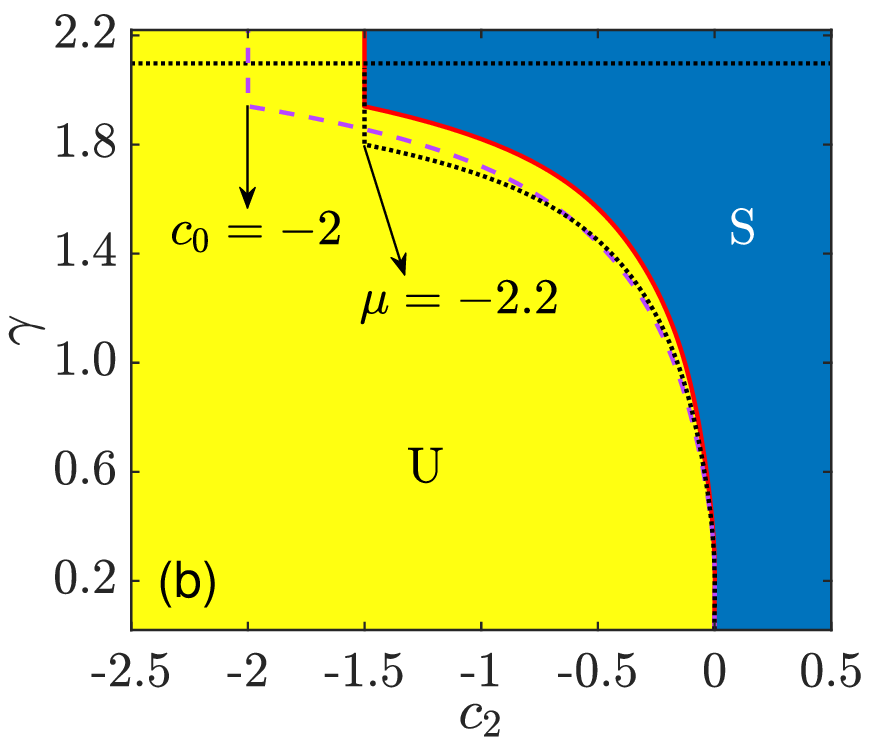}
	\end{minipage}}
	\caption{The stability domains of PW solitons (a) and SW solitons (b) about $c_2$ and $\gamma$. The blue (yellow) areas are stable (unstable) areas. PW solitons can not exist in white areas. Here, $\mu=-2.5$ and $c_0=-1.5$. The red solid line in (b) is the boundary between stable and unstable areas. Besides, the purple dashed line and the black dotted line represent the boundary when $c_0=-2$ and $\mu=-2.2$, respectively. In particular, the upper boundary of SW solitons becomes $\gamma=\sqrt{4.4}$ (the black horizontal line) when $\mu=-2.2$.}\label{fig3}
\end{figure}

With the exact solutions of PW1, PW2, SW1, and SW2 solitons, we can conveniently analyze their existence and stability domains. As we know, the attractive interaction is a necessary condition for the formation of bright solitons. Therefore, $c_0+c_2<0$ and $c_0<0$ are the necessary conditions to form PW and SW solitons, respectively. In addition, $k_1$ must be a real number, so $\mu<-\gamma^2/2$. For the stability of stationary solitons $\varphi_j(x)$, there are two methods to analyze it, one is the evolution of the stationary solitons with small perturbations, and the other is the linear stability analysis. The latter is more suitable for analyzing the stability of a large number of stationary solitons. We consider adding small perturbations to the stationary solutions $\varphi_j(x)$ as follows:
\begin{eqnarray}\label{eqrd}
\psi_{j}(x,t)=\left[\varphi_{j}(x)+a_{j}(x)\mathrm{e}^{\lambda t}+b_{j}^*(x)\mathrm{e}^{\lambda^* t}\right]\mathrm{e}^{-\mathrm{i} \mu t},
\end{eqnarray}
where $a_j$ and $b_j$ ($j=\pm1,0$) are perturbation functions, $\lambda$ is the growth rate of the perturbations, and '$\ast$' denotes complex conjugation. Substituting the perturbed solutions into equation~\eqref{eq1} and linearizing, we obtain an eigenvalue equation about perturbation functions $\mathrm{i}\boldsymbol L\boldsymbol\xi=\lambda \boldsymbol\xi$, where $\boldsymbol\xi=[a_{+1},b_{+1},a_{0},b_{0},a_{-1},b_{-1}]^T$, and $T$ represents transposition. $\boldsymbol L$ is a matrix, and its matrix elements are given in \ref{Appendix:A}. By numerical calculations, we can obtain the eigenvalues of matrix $\mathrm{i}\boldsymbol L$. When the real part maximum of $\lambda$ is greater than $10^{-3}$, we think that the stationary solutions are unstable because of the exponential growth of perturbations with time. Based on the linear stability analysis, we analyze the stability of PW and SW solitons in their existence domains, as shown in figure~\ref{fig3}.
We find that the stability domains of PW1 and PW2 solitons are the same, as well as for SW1 and SW2 solitons. For PW solitons, they are all stable in the existence domain ($c_2<-c_0$ and $\gamma<\sqrt{-2\mu}$), as shown in figure~\ref{fig3a}. However, not all SW solitons are stable. Interestingly, $c_2$ does not affect the expressions of SW solitons solutions but affects their stability, as shown in figure~\ref{fig3b}. According to the changes of the boundary between stable and unstable areas for different $c_0$ and $\mu$, we summarize some laws about the stability of SW solitons. When $c_2>0$ ($c_2<c_0$), SW solitons are stable (unstable). In addition, there are stable and unstable SW solitons in the area of $c_0<c_2<0$, which is divided by a curve. The curve intersects with the straight line $c_2=c_0$. We find the abscissa and ordinate of the intersection point depend on $c_0$ and $\mu$, respectively. In short, SW (PW) solitons exist stably in all antiferromagnetic (ferromagnetic) and partial ferromagnetic (antiferromagnetic) areas. Decreasing the interaction strength $c_0$ results in a larger stable ferromagnetic (antiferromagnetic) area.

\begin{figure}[h]
	\flushright
	\subfigure{\label{fig4a}
		\begin{minipage}[t]{0.35\linewidth}
			\includegraphics[width=5cm]{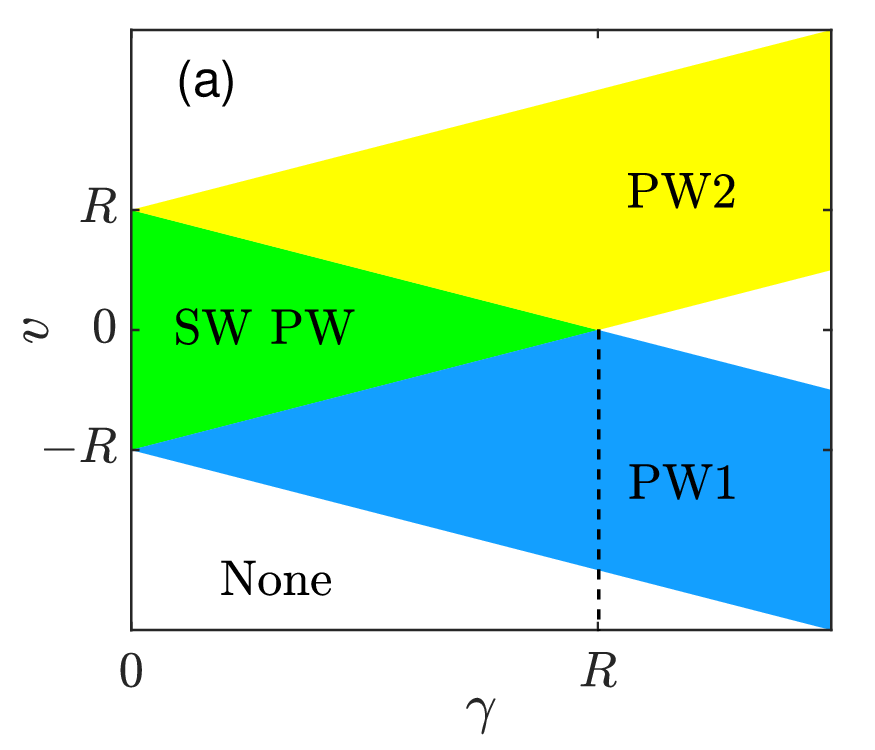}
	\end{minipage}}
	\subfigure{\label{fig4b}
		\begin{minipage}[t]{0.35\linewidth}
			\includegraphics[width=5cm]{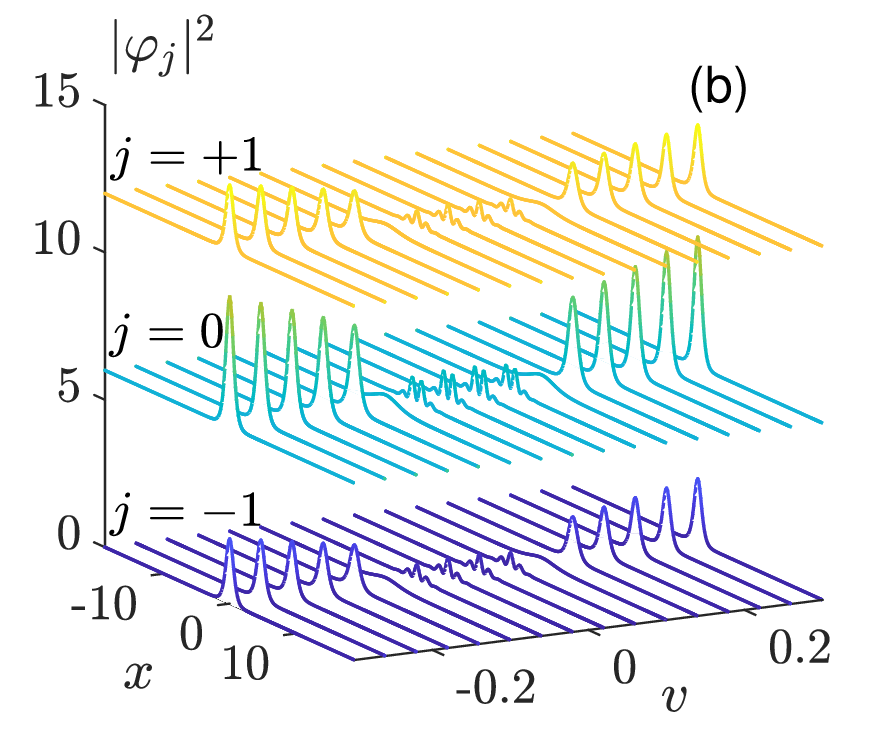}
	\end{minipage}}
	\subfigure{\label{fig4c}
		\begin{minipage}[t]{0.35\linewidth}
			\includegraphics[width=5cm]{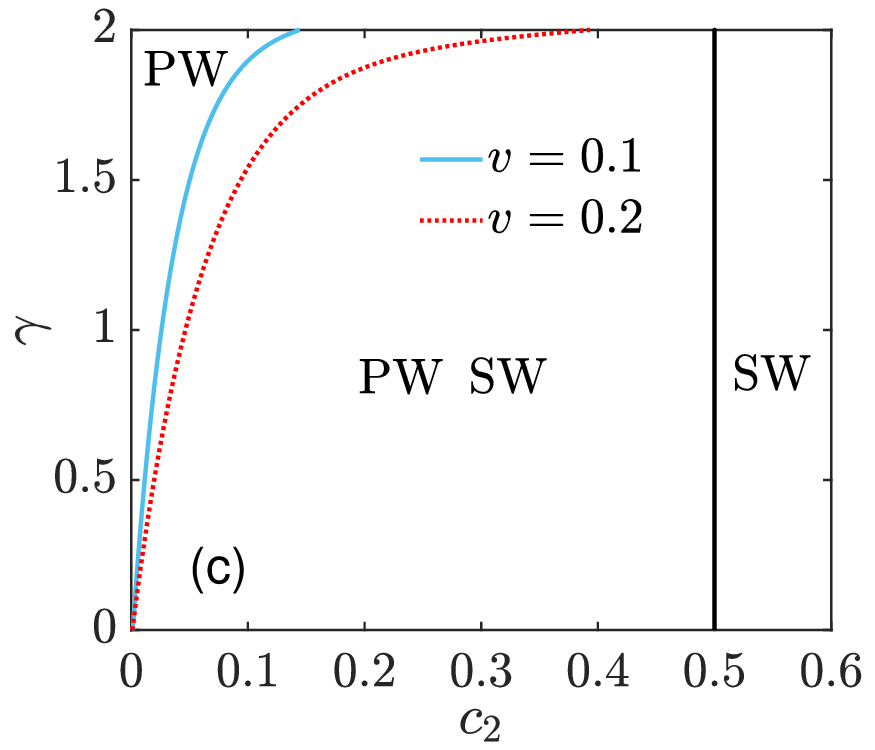}
	\end{minipage}}
	\subfigure{\label{fig4d}
		\begin{minipage}[t]{0.35\linewidth}
			\includegraphics[width=5cm]{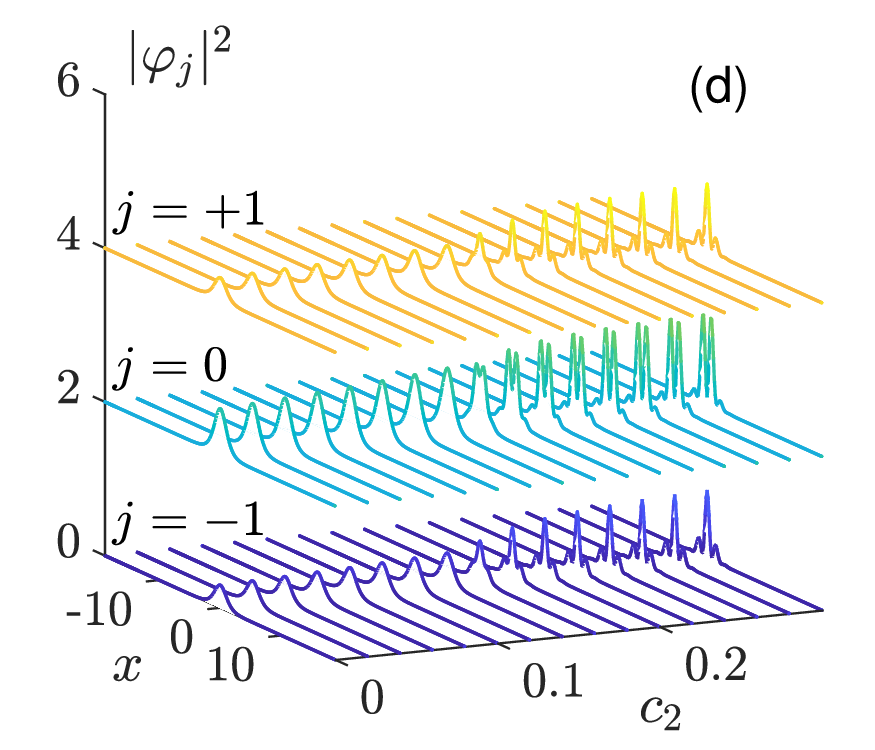}
	\end{minipage}}
	\caption{(a) The existence domains of moving PW and SW solitons about $\gamma$ and $v$. The bule (yellow) area is the existence domain of PW1 (PW2) solitons. All four types of bright solitons exist in the green area, but not in the white areas. Here, $R=\sqrt{-2\mu}$. (b) The three-component densities when $\mu=-2.5$, $\gamma=2.1$, $c_0=-0.5$, $c_2=0.3$, and $-0.3<v<0.3$. (c) The existence domains of moving PW and SW solitons about $\gamma$ and $c_2$ when $\mu=-2.5$ and $c_0=-0.5$. The blue solid (red dotted) line represents the dividing line when $v=0.1$ ($v=0.2$). The black solid line is the vanishing boundary of PW solitons. (d) The three-component densities corresponding to (c) for different $c_2$ when $\gamma=2$ and $v=0.1$. $|\varphi_{+1}|^2$ and $|\varphi_{0}|^2$ shown in (b) and (d) are translated.}\label{fig4}
\end{figure}

Next we discuss the moving solitons ($v\neq0$) which are stationary solutions in the moving frame. Although the Galilean invariance of equation~\eqref{eq1} is broken by the SOC effect, we can get the moving solitons by coordinate transformations
\begin{eqnarray}\label{eq7}
\begin{aligned}
&x^{\prime}=x-vt,~~t^{\prime}=t, \\
&\frac{\partial}{\partial x}=\frac{\partial}{\partial x^{\prime}},~~ \frac{\partial}{\partial t}=\frac{\partial}{\partial t^{\prime}}-v\frac{\partial}{\partial x^{\prime}}.
\end{aligned}
\end{eqnarray}
Substituting equation~\eqref{eq7} into equation~\eqref{eq1}, we obtain the equations in the moving frame. The details are in \ref{Appendix:A}. In the moving frame, the lowest branch becomes
\begin{eqnarray}\label{eq8}
E_{-}=\frac{k^2}{2}-vk-\sqrt{\gamma^2 k^2}.
\end{eqnarray}
We find $v$ breaks the axial symmetry of energy spectrum, and the existence domains of PW and SW solitons are also changed. For PW solitons, the exact solutions obtained by equations~\eqref{eq5a} and \eqref{eq6} are still the stationary solutions $\varphi_j(x^{\prime})\mathrm{exp}(-\mathrm{i}\mu t^{\prime})$ in the moving frame. Similarly, we gain the chemical potentials $\mu_{1,2}=-\left(\gamma\pm v\right)^2/2-k_1^2/2 $ corresponding to PW1 and PW2 solitons. When $\mu$ is fixed, the velocities of PW solitons are limited, the velocity ranges of PW1 and PW2 solitons are $-\sqrt{-2\mu}\mp\gamma<v<\sqrt{-2\mu}\mp\gamma$, respectively. As shown in figure~\ref{fig4a}, the plane $(\gamma,v)$ is divided by four rays which are determined by $\mu$.
PW1 and PW2 solitons exist in the blue and yellow areas, respectively. The green area is the common area of the blue and yellow areas, where both PW1 and PW2 solitons exist. For SW solitons, the exact solutions obtained by equations~\eqref{eq5b} and \eqref{eq6} do not represent moving SW solitons. We need to rely on numerical calculations (the squared-operator iteration method). We find SW solitons only exist in the green area, and the amplitudes of the SW solitons near the boundary of the green area are close to zero. Additionally, the limited velocity range of SW solitons is $-\sqrt{-2\mu}+\gamma<v<\sqrt{-2\mu}-\gamma$, which decreases as the increase of $\gamma$. Taking SW1 solitons as an example, selecting parameters $\mu=-2.5$, $c_0=-0.5$, and $c_2=0.3$, we show the three component densities when $\gamma=2.1$ in figure~\ref{fig4b}. We observe that there are only PW solitons and no SW solitons when the absolute value of velocity is larger than $\sqrt{5}-2.1$.

In addition to $\gamma$ and $v$, $c_0$ and $c_2$ are also important parameters affecting the existence domains of moving solitons, as shown in figure~\ref{fig4c}. For PW solitons, $c_2+c_0<0$ is still the existence condition of moving PW solitons. But for SW solitons, $c_2$ can not be taken arbitrarily. SW solitons are limited to the right area of the blue solid (red dotted) curve on the plane ($c_2,\gamma$) when $v=0.1$ ($v=0.2$). For example, noting the profiles of SW1 solitons when $\mu=-2.5$, $\gamma=2$, $c_0=-0.5$, and $v=0.1$, as shown in figure~\ref{fig4d}. We find the stripe contrast of the SW1 solitons becomes smaller as $c_2$ decreases, and the phase transition occurs when $c_2\approx0.155$. In a word, moving PW and SW solitons have a common area surrounded by the straight line $c_2=-c_0$ and a curve determined by $v$. In addition, moving PW and SW solitons also exist on the left and right of the common area, respectively. When $v<0$, the existence domain of SW solitons is consistent with that of $|v|$.

By discussing the stability of moving solitons in the existence domains, we find that the moving PW and SW solitons are all stable. Comparing figure~\ref{fig4c} with figures~\ref{fig3a} and \ref{fig3b}, this is understandable because the moving solitons only exist in the stable area of the stationary solitons.

\subsection{With a Zeeman field}
For the case that the Zeeman field exists ($p\neq0$), $\psi_{+1}$ and $\psi_{-1}$ in equation~\eqref{eq1} are no longer equivalent. It is difficult to obtain exact analytical solutions, so we use the numerical method to find PW and SW solitons of equation~\eqref{eq1}, including stationary and moving solitons. In addition, we only consider the case of $p>0$ because it is symmetrical with the case of $p<0$.

\begin{figure}[h]
	\flushright
	\subfigure{\label{fig5a}
		\begin{minipage}[t]{0.31\linewidth}
			\includegraphics[width=5cm]{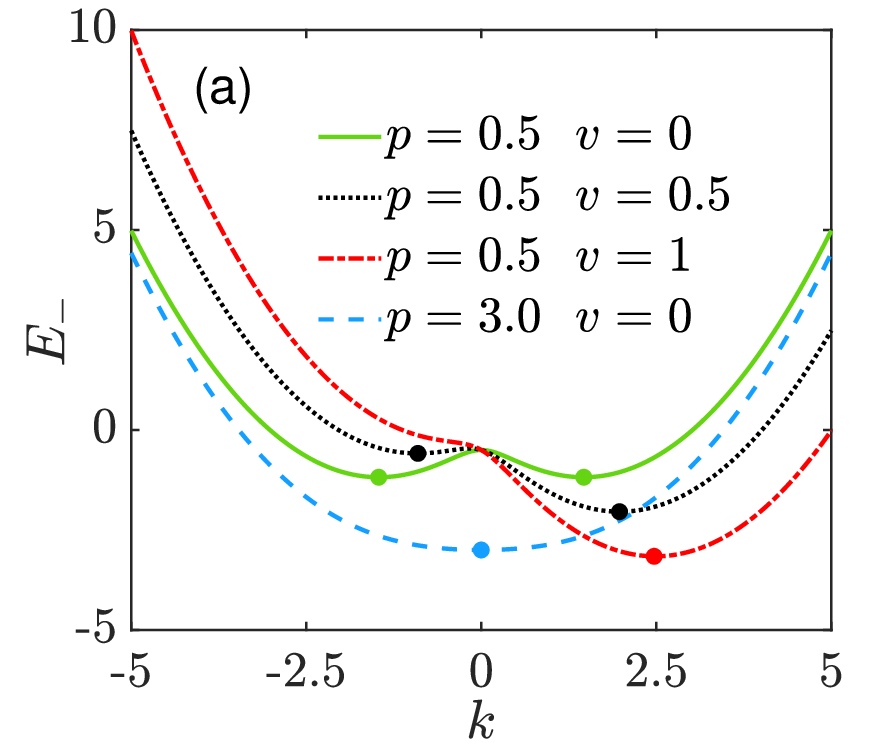}
	\end{minipage}}
	\subfigure{\label{fig5b}
		\begin{minipage}[t]{0.31\linewidth}
			\includegraphics[width=5cm]{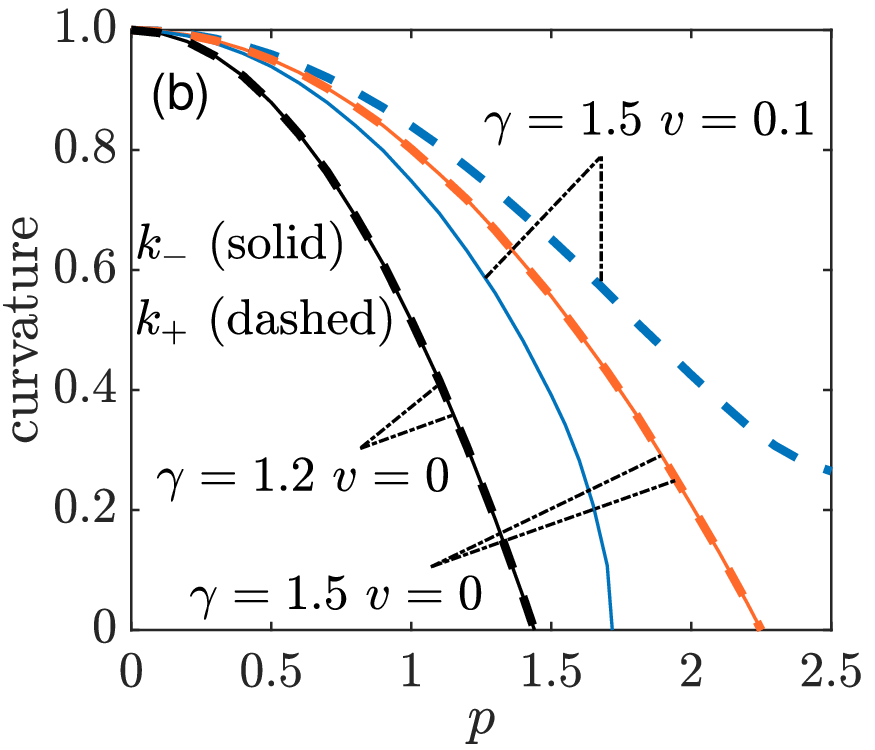}
	\end{minipage}}
	\subfigure{\label{fig5c}
		\begin{minipage}[t]{0.31\linewidth}
			\includegraphics[width=5cm]{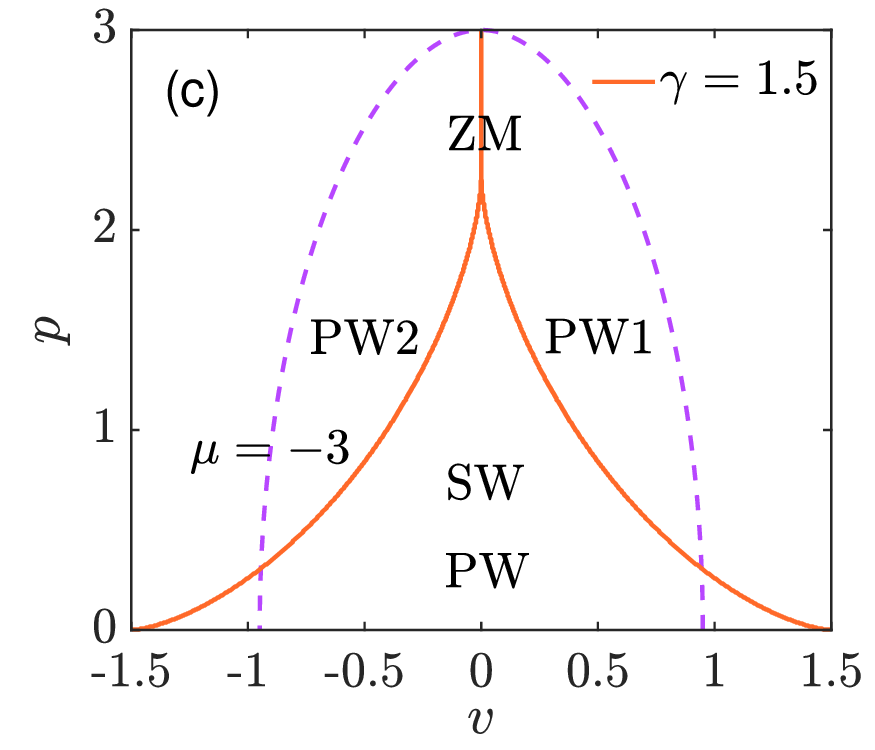}
	\end{minipage}}
	\caption{(a) The lowest branch $E_-$ of energy spectrum for different $v$ and $p$. The points represent the local minimums. Here, $\gamma=1.5$. (b) The curvatures at the local minimum points in different situations. (c) The existence domains of different types of solitons divided by the orange solid lines on the plane $(v,p)$ when $\gamma=1.5$. The purple dotted line represents the contour line with $\min (E_-)=-3$. The existence condition of bright solitons is $\mu<\min (E_-)$.}
\end{figure}

We still analyze from the perspective of the energy spectrum. In the moving frame, the lowest branch becomes
\begin{eqnarray}\label{eq9}
E_{-}=\frac{k^2}{2}-vk-\sqrt{\gamma^2 k^2+p^2}.
\end{eqnarray}
$v$, $\gamma$ and $p$ affect the number of local minimums of $E_-$, which determines the types of solitons in the system. We know that $\gamma$ is the only reason why $E_-$ has a double-valley structure. The structure of $E_-$ and the curvatures at the local minimum points are depicted in figures~\ref{fig5a} and \ref{fig5b}.
We find that the increase of $p$ causes $k_+$ and $k_-$ to approach each other, and the two local minimums and their corresponding curvatures both decrease. $v$ makes one of the local minimums (the two corresponding curvatures) increase and the other decrease. Whether $v$ or $p$ become large enough, the double-valley structure of $E_-$ is broken. In this case, there are only PW solitons and no SW solitons in the system. Basing on the number of local minimums of $E_-$, we obtain the existence domains of different types of solitons. As shown in figure~\ref{fig5c}, the orange solid lines determined by $\gamma$ divide the plane ($v,p$) into three areas when $\gamma=1.5$. ZM solitons only exist in the case of $p>\gamma^2$ and $v=0$, except in this case, there are always PW solitons present. SW solitons exist in the area with small absolute values of $p$ and $v$. When $\mu$ is fixed, these bright solitons only exist in the semi-ellipse area surrounded by the purple dashed line. The larger $\mu$, the smaller the semi-ellipse area. Next, we mainly discuss PW and SW solitons.

\begin{figure}[h]
	\flushright
	\subfigure{\label{fig6a}
		\begin{minipage}[t]{0.35\linewidth}
			\includegraphics[width=5cm]{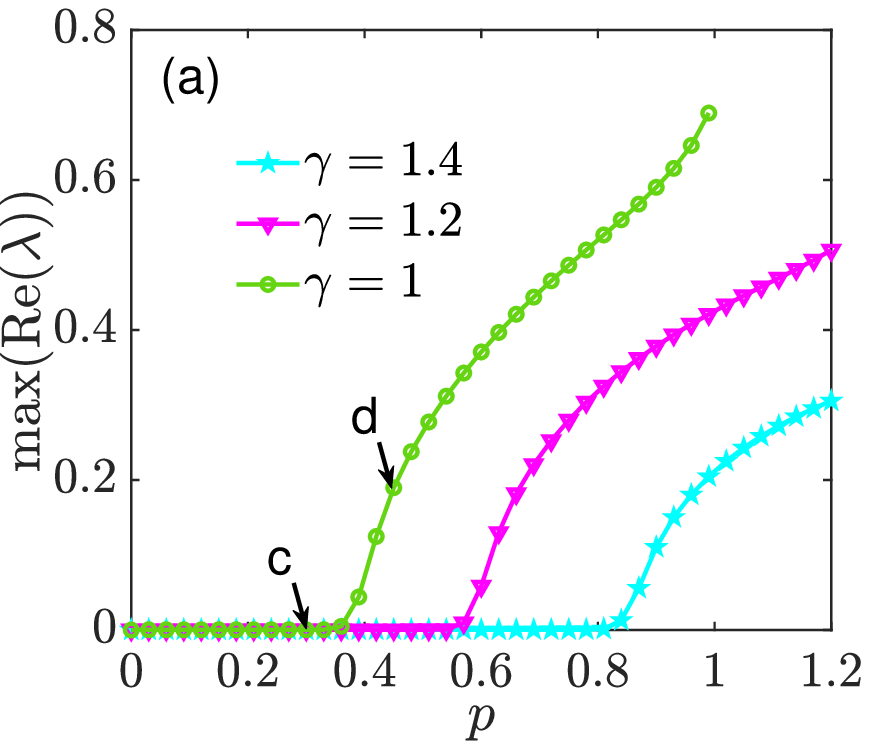}
	\end{minipage}}
	\subfigure{\label{fig6b}
		\begin{minipage}[t]{0.35\linewidth}
			\includegraphics[width=5cm]{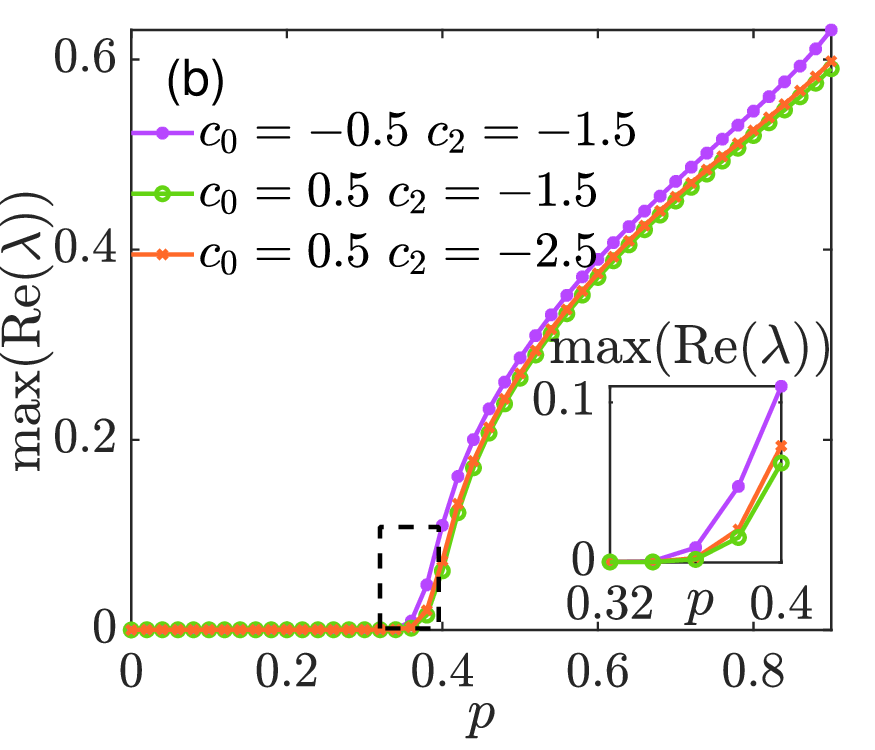}
	\end{minipage}}
	\subfigure{\label{fig6c}
		\begin{minipage}[t]{0.35\linewidth}
			\includegraphics[width=5cm]{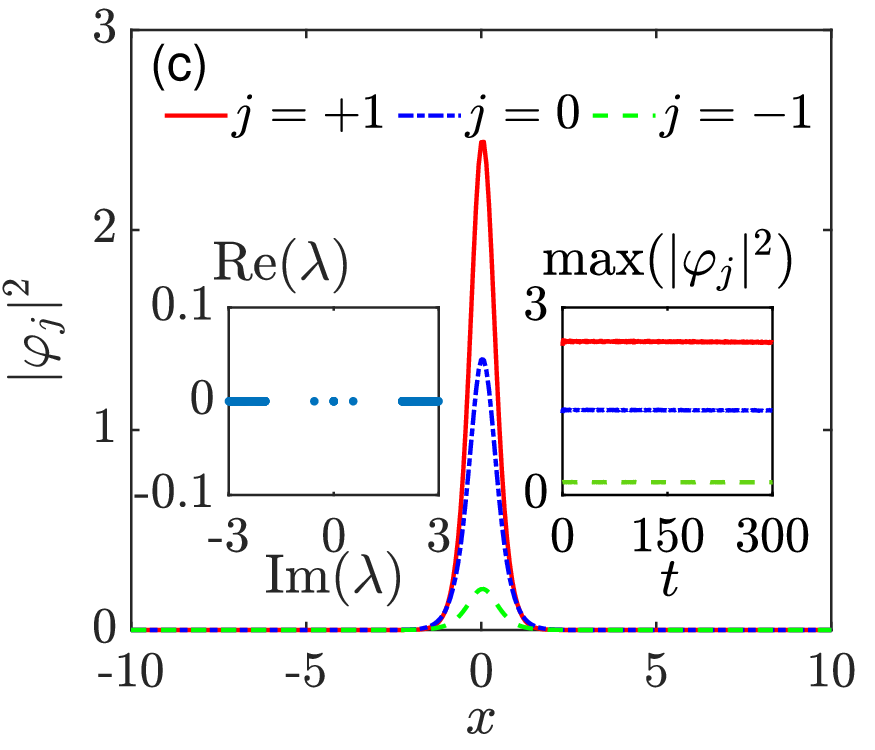}
	\end{minipage}}
	\subfigure{\label{fig6d}
		\begin{minipage}[t]{0.35\linewidth}
			\includegraphics[width=5cm]{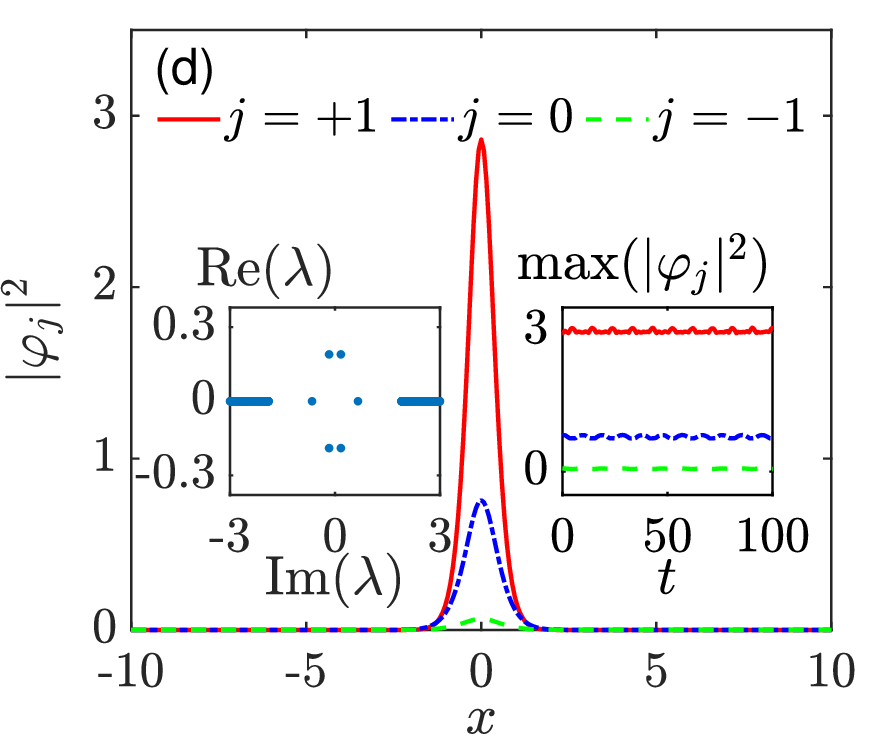}
	\end{minipage}}
	\caption{The influence of $p$ on the stability of PW solitons for different $\gamma$ (a) or for different $c_0$ and $c_2$ (b). (a) $c_0=0.5$ and $c_2=-1.5$. (b) $\gamma=1$. The subgraph in (b) is an enlarged view of the dashed frame. Here, $\mu=-2.5$ and $v=0$. (c) and (d) show the PW solitons in (a) when $p=0.3$ and $p=0.45$, respectively. Besides, the left and right subgraphs represent the linear stability analysis and the evolution of $\max(|\varphi_j|^2)$ under 2\% random-noise perturbations, respectively.}
\end{figure}

Firstly, we obtain PW solitons in the existence domains and discuss their stability by numerical calculations. The necessary condition is $c_0+c_2<0$. When $v=0$, the stability domains of PW1 and PW2 solitons are completely consistent because of the symmetrical energy spectrum. By the control variable method, we systematically investigate the influence of various parameters on the stability of PW solitons, as shown in figures~\ref{fig6a} and \ref{fig6b}.
We find a commonality that the stable and unstable areas are separated by a critical value $p_0$. When $p$ is greater (small) than $p_0$, the corresponding PW soliton is unstable (stable). $p_0$ becomes larger with increasing $\gamma$. However, in the case of $c_2<0$, $p_0$ hardly changes when we adjust $c_0$ and $c_2$. It indicates that the stability of PW solitons is related to the single-particle energy spectrum. In addition, we give two examples to illustrate the correctness of the linear stability analysis, as shown in figures~\ref{fig6c} and \ref{fig6d}. We find $p$ makes the proportion of the three component densities change and affects the stability. During the real-time evolution of solitons with initial perturbations, the amplitudes of stable ($p=0.3$) and unstable ($p=0.45$) PW solitons remain unchanged and oscillate periodically, respectively, which is consistent with the results of the linear stability analysis.

\begin{figure}[h]
	\flushright
	\subfigure{\label{fig7a}
		\begin{minipage}[t]{0.35\linewidth}
			\includegraphics[width=5cm]{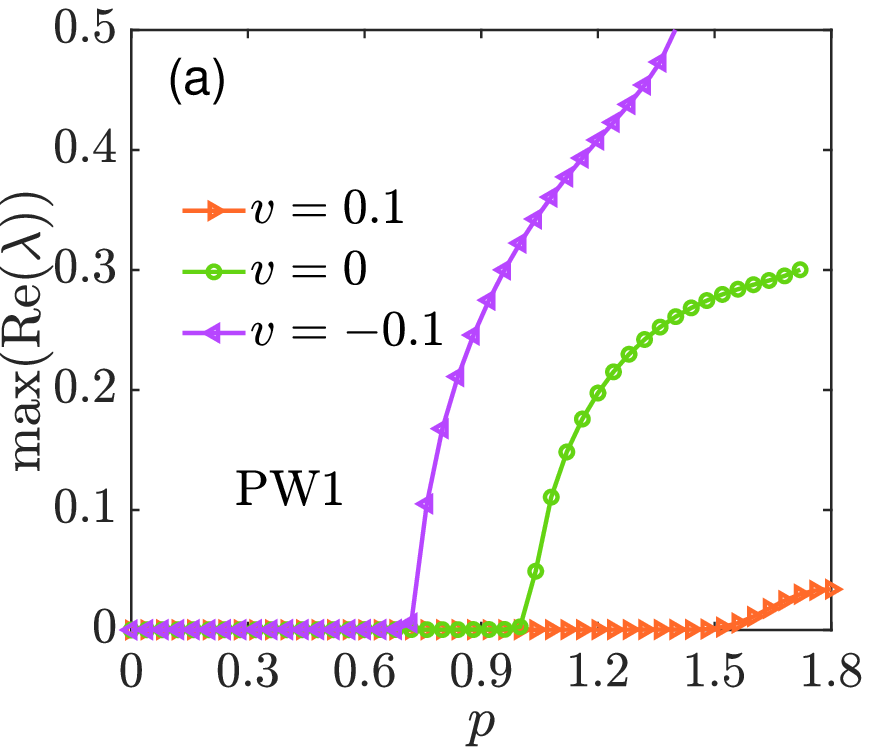}
	\end{minipage}}
	\subfigure{\label{fig7b}
		\begin{minipage}[t]{0.35\linewidth}
			\includegraphics[width=5cm]{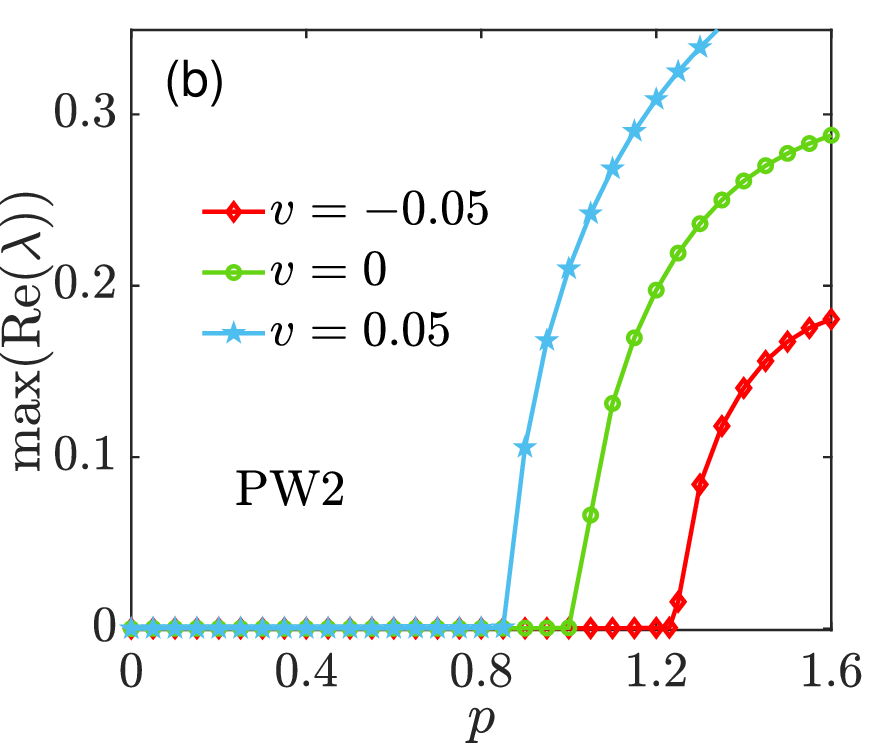}
	\end{minipage}}
	\caption{The influence of $p$ on the stability of PW1 solitons (a) and PW2 solitons (b) for different $v$. Here, $\mu=-2.5$, $\gamma=1.5$, $c_0=0.5$, and $c_2=-1.5$.}
\end{figure}

We also discuss the influence of velocity on the stability of PW1 and PW2 solitons. As shown in figures~\ref{fig7a} and \ref{fig7b}, we find an interesting phenomenon that $v$ has opposite influence on the stability of PW1 and PW2 solitons.
In other words, the critical values $p_0$ of the PW1 solitons ($v>0$) and the PW2 solitons ($v<0$) are all larger than that of the PW solitons ($v=0$). Besides, we find that the promoting effect of $v$ on stability is stronger than the inhibition effect. Combining the influence of $\gamma$, $c_0$, and $c_2$ on stability, we find the parameters which change the structure of $E_-$ greatly affect the stability of PW solitons, such as $\gamma$, $p$, and $v$. Further, the parameter that increase the curvature of the local minimum point can improve the stability of the corresponding PW soliton. $c_0$ and $c_2$, which do not change the curvature, barely affect the stability of PW solitons.

\begin{figure}[h]
	\flushright
	\subfigure{\label{fig8a}
		\begin{minipage}[t]{0.35\linewidth}
			\includegraphics[width=5cm]{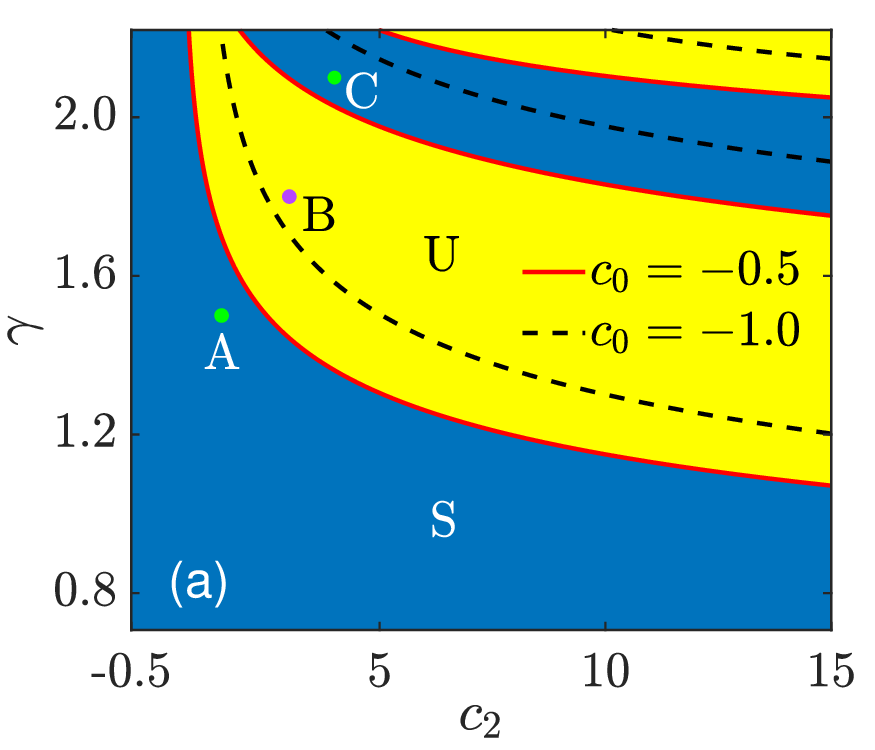}
	\end{minipage}}
	\subfigure{\label{fig8b}
		\begin{minipage}[t]{0.35\linewidth}
			\includegraphics[width=5cm]{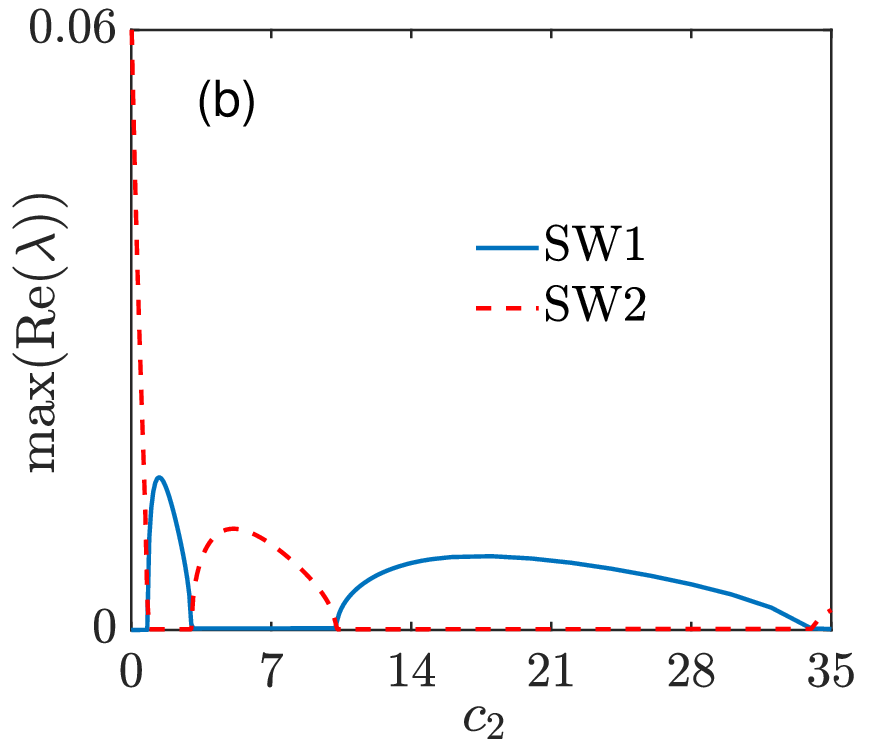}
	\end{minipage}}
	\caption{(a) The stability domains of SW1 solitons on the plane ($c_2, \gamma$) for different $c_0$. The blue and yellow areas are stable and unstable areas of SW1 solitons, respectively, but for SW2 solitons, the opposite is true. The red solid lines are the dividing lines when $c_0=-0.5$. Besides, the dividing lines move to the black dashed lines when $c_0=-1$. The coordinates of three markers $\mathrm{A}$, $\mathrm{B}$, and $\mathrm{C}$ are $(1.5,1.5)$, $(3,1.8)$, and $(4,2.1)$, respectively. (b) The real part maximums of $\lambda$ for different $c_2$ when $\gamma=2.1$ and $c_0=-0.5$. The blue solid line and the red dashed line correspond to SW1 and SW2 solitons, respectively. Here, $\mu=-2.5$ and $p=0.5$.}
\end{figure}

For the stability of SW solitons, we consider the case of $v=0$ firstly. In addition to the condition $p<\gamma^2$, $c_0<0$ is also a premise for the existence of SW solitons. From figure~\ref{fig3b}, we know that SW solitons are unstable in the absence of Zeeman field when $c_2<c_0$. Therefore, we focus on discussing the stability of SW solitons for $c_2>c_0$ when $p\neq0$. In figure~\ref{fig8a}, the stable (blue) and unstable (yellow) areas of SW1 solitons are arranged alternately.
When adjusting $c_0$ from $-0.5$ to $-1$, the dividing lines between stable and unstable areas change from the red solid lines to the black dashed lines. Moreover, we find that SW1 solitons are all stable when $c_0<c_2<0$, which is caused by $p\neq0$. It is different from the situation without a Zeeman field. The stability of SW2 solitons on the plane ($c_2,\gamma$) is also analyzed. Coincidentally, the stability domains of SW1 and SW2 solitons are complementary. In other words, SW2 solitons are unstable in the stable areas of SW1 solitons. As shown in figure~\ref{fig8b}, we can clearly understand the complementarity, by observing the change of the real part maximum of $\lambda$ when $c_2$ increases. In addition, the alternating changes of the stability domains are not just twice. SW1 (SW2) solitons are stable (unstable) again when $c_2=34$. Thus, when $c_2$ continues to increase, there are still alternate stable and unstable areas. The instability of the unstable areas gradually weaken.

\begin{figure}[h]
	\flushright
	\includegraphics[width=0.75\linewidth]{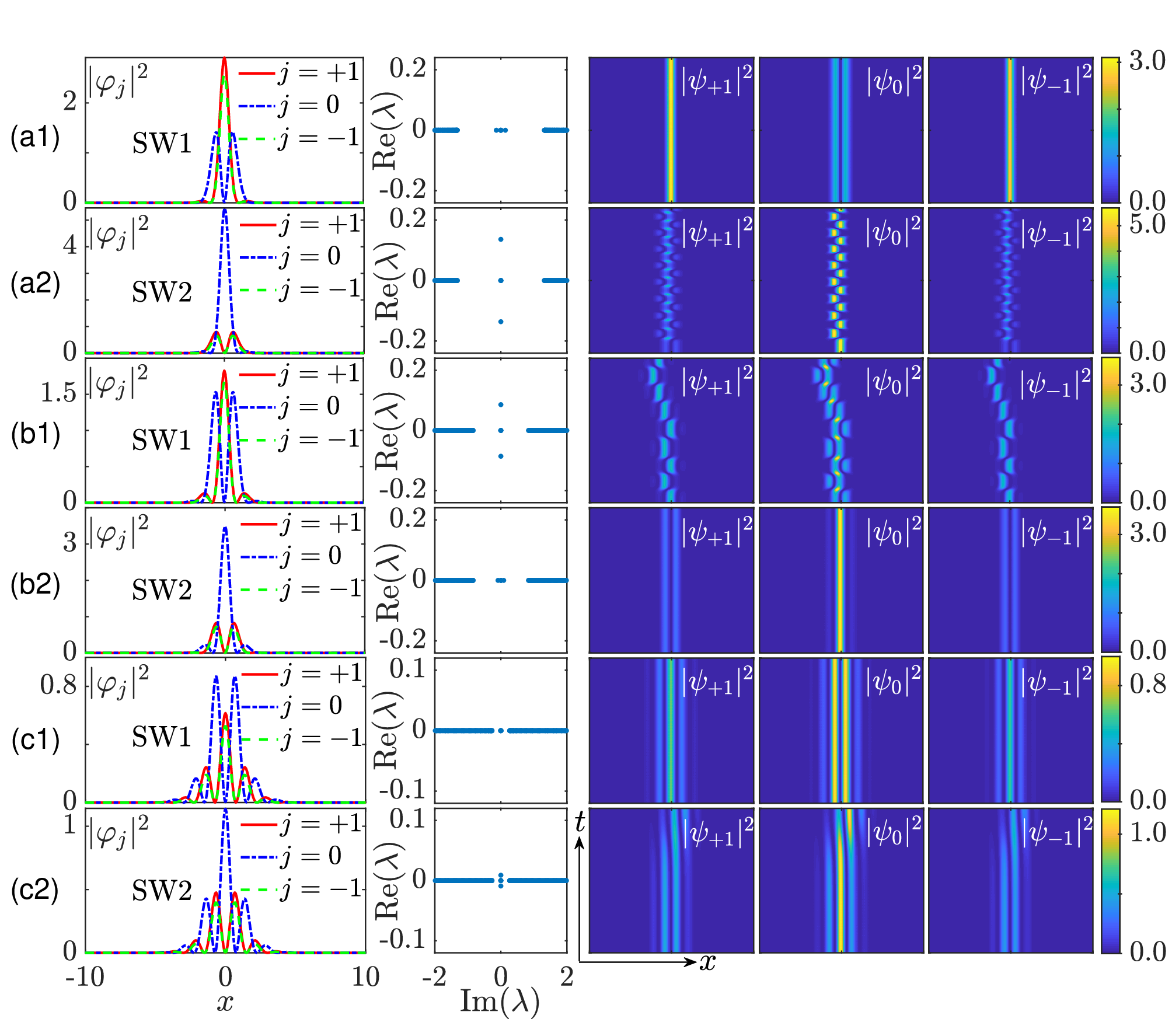}
	\caption{The profiles, stability, and evolution of the SW1 (odd rows), and SW2 solitons (even rows). The parameters of (a1) and (a2), (b1) and (b2), (c1) and (c2) correspond to $\mathrm{A}$, $\mathrm{B}$, $\mathrm{C}$ in figure~\ref{fig8a}, respectively. The first and second columns on the left represent the profile and stability of SW solitons, respectively. The three columns on the right represent the evolution of the three components under 2\% random-noise perturbations, their transverse windows $x\in\left[-10,10\right]$ and longitudinal windows $t\in\left[0,1000\right]$. Besides, the colorbars for each row are shared.}
	\label{fig9}
\end{figure}

Selecting the parameters corresponding to $\mathrm{A}$, $\mathrm{B}$, and $\mathrm{C}$ in figure~\ref{fig8a}, we draw the profiles and stability of the corresponding SW1 and SW2 solitons, and show their evolutions, as shown in figure~\ref{fig9}.
We find the stability of the SW1 and SW2 solitons is indeed opposite, and the SW solitons with lower unstable growth rates can remain stable for longer time. However, the unstable SW solitons do not immediately disperse after deformation, they begin to change their profiles regularly. The reason is that the nonlinear terms in the Hamiltonian of equation~\eqref{eq1} are equivalent to a time-dependent external potential when the SW solitons are unstable. The time-dependent external potential cause the quantum transport phenomena \cite{fu_nonlinear_2022}, which is similar to the unstable evolution in figure~\ref{fig9}.

\begin{figure}[h]
	\flushright
	\subfigure{\label{fig10a}
		\begin{minipage}[t]{0.35\linewidth}
			\includegraphics[width=5cm]{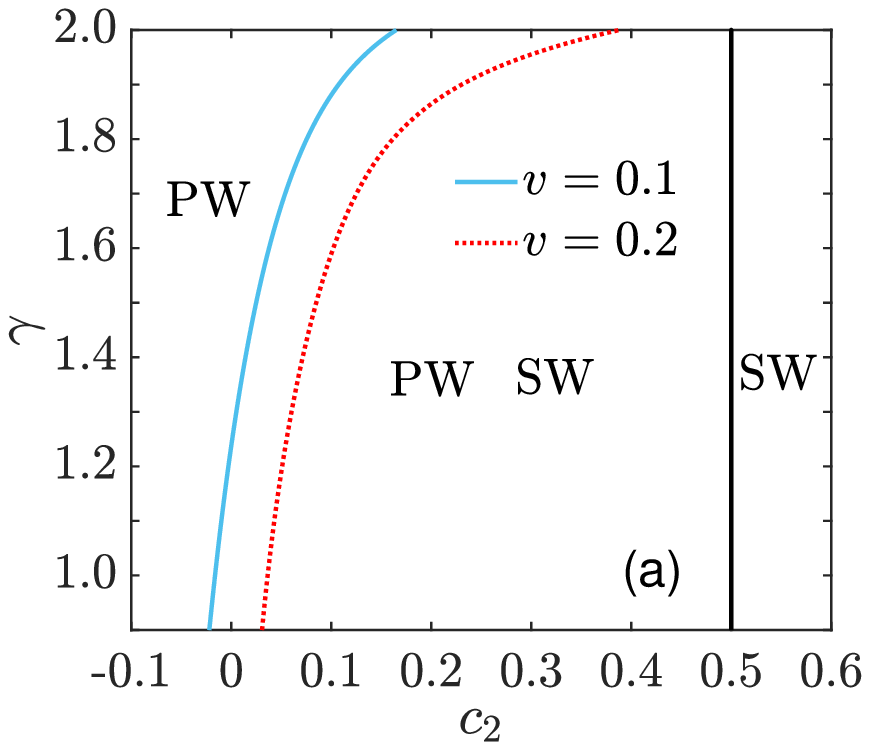}
	\end{minipage}}
	\subfigure{\label{fig10b}
		\begin{minipage}[t]{0.35\linewidth}
			\includegraphics[width=5cm]{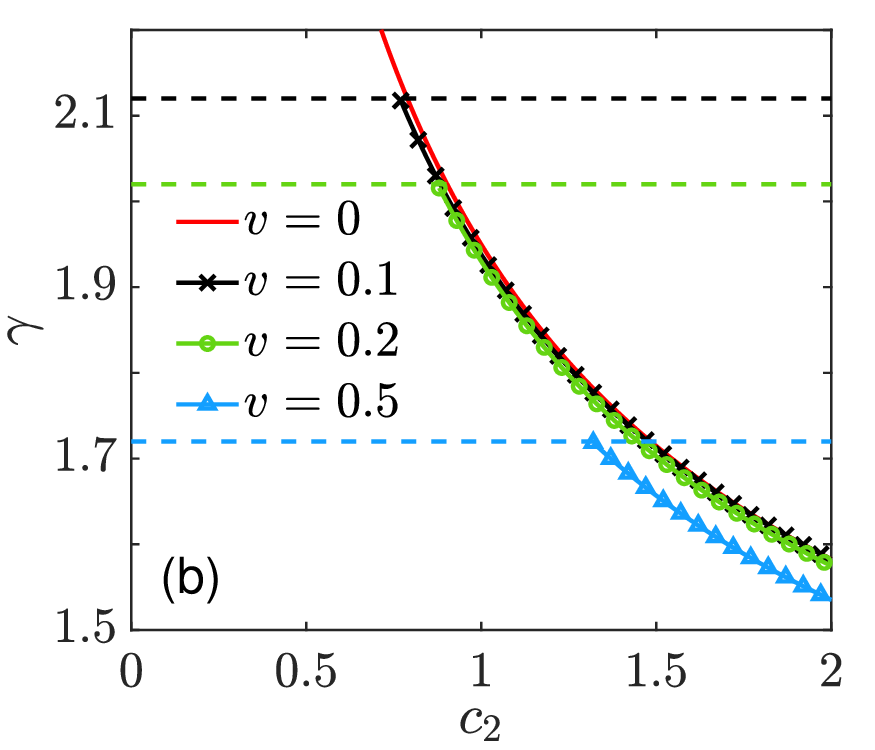}
	\end{minipage}}
	\caption{(a) The existence domains of moving PW and SW solitons for different $v$ when $p=0.5$. The blue solid (red dotted) line represents the dividing line when $v=0.1$ ($v=0.2$). The black solid line is the vanishing boundary of PW solitons. Here, $\mu=-2.5$ and $c_0=-0.5$. (b) The changes of the leftmost dividing line in figure~\ref{fig8a} when $v\neq0$. The dashed lines represent the upper boundaries of the existence domains of SW solitons at different velocities.}
\end{figure}

When $v\neq0$, SW solitons are limited to the right area of the blue (red) curve in figure~\ref{fig10a} when $v=0.1$ ($v=0.2$).
Different from the situation without the Zeeman field, the linear Zeeman effect $p$ relax this restriction. Comparing with figure~\ref{fig4c}, we find the existence domain of moving SW solitons expands slightly to the right when $p\neq0$, and even contains a small portion of ferromagnetic area ($c_2<0$) when $v=0.1$. For the stability of moving SW solitons, $v$ only causes slight movement of the dividing lines of stability domain, especially when $v$ is small. As shown in figure~\ref{fig10b}, we only show the changes of the leftmost dividing line in figure~\ref{fig8a} to observe clearly.

\section{Collision dynamics of stable solitons}\label{sec5}
The existence of stable moving solitons raises the question about their
interactions. As we know, when the relative velocity of two solitons is large, interference fringes are generated in the collision area \cite{helm_sagnac_2015}. Here, we only discuss the collisions between moving solitons with small velocities, and concentrate on the dynamics of the total particle density $n$ and $z$-component spin density $F_z=|\psi_{+1}|^2-|\psi_{-1}|^2$.

\begin{figure}[h]
	\flushright
	\includegraphics[width=0.75\linewidth]{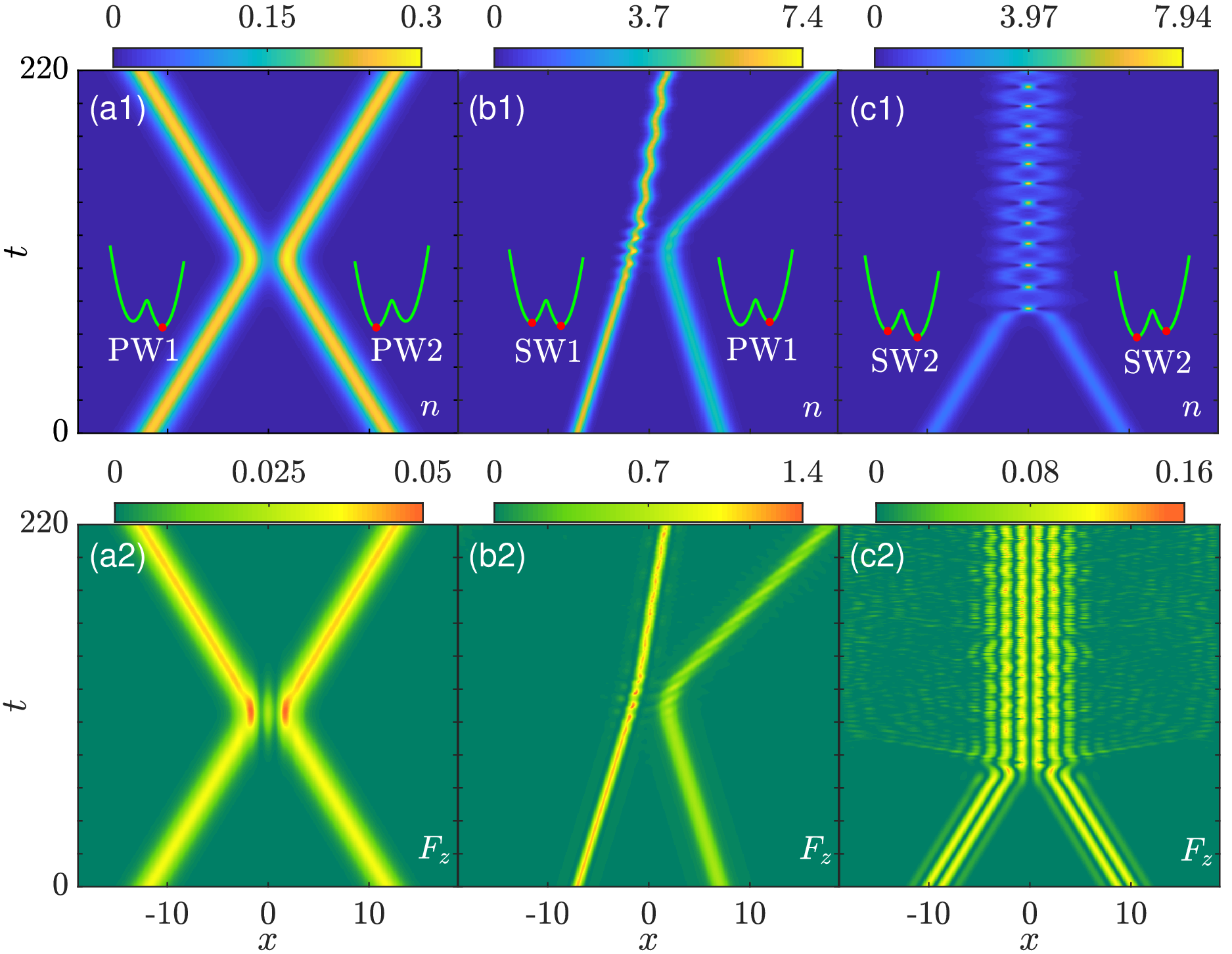}
	\caption{The collisions between different types of solitons. (a1-c1) and (a2-c2) display the evolution of the total particle density $n$ and the $z$-component spin density $F_z$, respectively. The subgraphs show the lowest branch $E_-$ and the corresponding local minimums of solitons. Here, $\mu=-2.5$, $\gamma=2$, $c_0=-0.5$, and $p=0.5$, but for the SW1 soliton in (b1) $\mu=-3.5$. The other parameters are $c_2=-1.5$, $v=\pm0.1$ in (a1) and (a2), $c_2=0.2$, $v=\pm0.05$ in (b1) and (b2), $c_2=3$, $v=\pm0.1$ in (c1) and (c2).}
	\label{fig11}
\end{figure}

Selecting two solitons with opposite velocities in the stable areas and putting them in different positions, we find some interesting collision phenomena, as shown in figure~\ref{fig11}.
In the two left figures, the PW1 and PW2 solitons bounce apart from each other after colliding, which occupy the minimums of their corresponding $E_-$ respectively. There are no transferred particles between the two solitons, but particles transfer between three components, as can be seen from the increase of $F_z$ after colliding. In addition, the collision between the PW2 and PW1 solitons which occupy the larger local minimums makes $F_z$ decrease, not shown here. Then noting the collision between the SW1 and PW1 solitons with different chemical potential $\mu$ in the two middle figures, we find the macroscopic tunneling effects \cite{zhao_tunneling_2017,jia_soliton_2022} near the collision area. After colliding, the SW1 soliton begins to oscillate periodically, and the velocity of the PW1 soliton changes. The $z$-component spin densities $F_z$ of the two solitons also begin to oscillate slightly. Furthermore, the most interesting phenomenon occurs in the collision between the two SW2 solitons in the right figures. The two SW2 solitons attract each other as they approach, leading to quasi-periodic collisions, during which a small portion of particles scatter and detach from the main body. $F_z$ still maintains a relatively good stripe shape after colliding. After more discussion, we find this phenomenon can also occur in the collision between two SW1 solitons with equal chemical potential $\mu$.

\section{Conclusions}\label{sec6}
In summary, we investigated the stationary and moving bright solitons in spin-1 BECs with SOC in a Zeeman field. Combining the single-particle energy spectrum, exact solutions, and numerical solutions, we systematically analyzed the existence and stability domains of four types of bright solitons (PW1, PW2, SW1, and SW2 solitons). Moreover, we briefly discussed the collision dynamics of different types of stable solitons and found some interesting inelastic collision phenomena. We have summarized some laws about the existence and stability of PW and SW solitons.

For the existence of PW and SW solitons, we find that the single-particle energy spectrum contains a lot of useful information. According to its local minimums, we can obtain the range of the chemical potential $\mu$ corresponding to solitons. By discussing the number of the local minimums, we can accurately predict the type of solitons which exist in the system. PW solitons exist when the energy spectrum has one or more local minimums. SW solitons only exist when the energy spectrum has two local minimums. We further find the interactions $c_0$ and $c_2$ are also key factors affecting the existence of solitons, and the necessary conditions for the existence of PW and SW solitons are $c_0+c_2<0$ and $c_0<0$, respectively. In particular, for moving SW solitons, $c_2$ is restricted and must be greater than zero. Fortunately, the linear Zeeman effect $p$ can relax this restriction and make the moving SW solitons exist in a small portion of antiferromagnetic areas ($c_2<0$).

For the stability of PW and SW solitons in the existence domains, we summarize the influence of various parameters on stability. In the absence of the Zeeman field, all stationary and moving PW solitons in the existence domain are stable. Stationary SW solitons stably exist in the ferromagnetic area ($c_2>0$) and a small portion of the antiferromagnetic area ($c_2<0$), but moving SW solitons are only stable in the ferromagnetic area because of the limitation of $c_2$. When $p\neq0$, the stable areas vary greatly. For PW solitons, when $p>p_0$, the PW soliton is unstable. The parameters which change the curvature at the local minimum point significantly affect the stability of the corresponding PW soliton, such as $\gamma$, $p$, and $v$. Increasing the curvature at the local minimum point makes the corresponding PW soliton more stable. Other parameters barely affect the stability of PW solitons, which do not change the curvature, such as $c_0$ and $c_2$. For SW solitons, the linear Zeeman effect $p$ makes it impossible for both SW1 and SW2 solitons to exist stably under the same conditions. Their stable areas are complementary and staggered on the plane ($c_2,\gamma$). SW1 solitons are even stable in the ferromagnetic area where SW1 solitons are unstable when $p=0$. Adjusting $c_0$ by the Feshbach resonance technology, we can change the stability of SW solitons. Moreover, we find $v$ slightly affects the stability of SW solitons. The stable areas of the moving SW solitons with small velocities and the stationary SW solitons are almost the same.

Our results are helpful for theoretical and experimental work on PW and SW phases, and the properties of dark solitons in the system can also be explored by similar methods.

\ack{We acknowledge support of the National Natural Science Foundation of China, No. 11835011.}

\section*{Data availability statement}
The data that support the findings of this study are available upon reasonable request from the authors.

\appendix

\section{The equations in the moving frame and the linear stability analysis}\label{Appendix:A}
By coordinate transformations \eqref{eq7}, we obtain the equations describing a spin-orbit-coupled spin-1 BEC in the moving frame
\begin{eqnarray}
\begin{aligned}\label{eqA1}
\mathrm{i}\frac{\partial \psi_{\pm 1}}{\partial{t}}=&\left(-\frac{1}{2} \frac{\partial^{2}}{\partial{x}^{2}}+c_0n+c_2\left(n_{\pm 1}+n_0-n_{\mp 1}\right)\right)\psi_{\pm 1}\\&+c_{2} \psi_{0}^{2} \psi_{\mp 1}^{*}-\frac{\mathrm{i}\gamma}{\sqrt{2}}\frac{\partial\psi_0}{\partial x}+(q\mp p)\psi_{\pm 1}+\mathrm{i}v\frac{\partial \psi_{\pm 1}}{\partial x},\\
\mathrm{i}\frac{\partial \psi_{0}}{\partial{t}}=&\left(-\frac{1}{2} \frac{\partial^{2}}{\partial{x}^{2}}+c_0n+c_2\left(n_{+1}+n_{-1}\right)\right)\psi_{0}\\&+2c_{2} \psi_{0}^{*} \psi_{+1}\psi_{-1}-\frac{\mathrm{i}\gamma}{\sqrt{2}}\left(\frac{\partial\psi_{+1}}{\partial x}+\frac{\partial\psi_{-1}}{\partial x}\right)+\mathrm{i}v\frac{\partial \psi_{0}}{\partial x}.
\end{aligned}
\end{eqnarray}
Substituting $\psi_j(x,t)=\varphi_j(x)\mathrm{e}^{-\mathrm{i}\mu t}$ into equation~\eqref{eqA1}, we obtain the stationary state equations,
\begin{eqnarray}
\begin{aligned}\label{eqA2}
\mu \varphi_{\pm 1}=&\left(-\frac{1}{2} \frac{\partial^{2}}{\partial{x}^{2}}+c_0n+c_2\left(n_{\pm 1}+n_0-n_{\mp 1}\right)\right)\varphi_{\pm 1}\\&+c_{2} \varphi_{0}^{2} \varphi_{\mp 1}^{*}-\frac{\mathrm{i}\gamma}{\sqrt{2}}\frac{\partial\varphi_0}{\partial x}+(q\mp p)\varphi_{\pm 1}+\mathrm{i}v\frac{\partial \varphi_{\pm 1}}{\partial x},\\
\mu\varphi_{0}=&\left(-\frac{1}{2} \frac{\partial^{2}}{\partial{x}^{2}}+c_0n+c_2\left(n_{+1}+n_{-1}\right)\right)\varphi_{0}\\&+2c_{2} \varphi_{0}^{*} \varphi_{+1}\varphi_{-1}-\frac{\mathrm{i}\gamma}{\sqrt{2}}\left(\frac{\partial\varphi_{+1}}{\partial x}+\frac{\partial\varphi_{-1}}{\partial x}\right)+\mathrm{i}v\frac{\partial \varphi_{0}}{\partial x}.
\end{aligned}
\end{eqnarray}
Both the stationary and moving solitons can be obtained by solving equation~\eqref{eqA2} numerically when the exact solutions are difficult to be obtained.

The following describes the steps of the linear stability analysis. Substituting the perturbed solutions equation~\eqref{eqrd} into equation~\eqref{eqA1} and linearizing, we obtain the equations about the perturbation functions
\begin{eqnarray}
\mathrm{i}\boldsymbol L\boldsymbol \xi=\lambda \boldsymbol \xi,
\end{eqnarray}
where $\boldsymbol \xi=\left[a_{+1},b_{+1},a_{0},b_{0},a_{-1},b_{-1}\right]^T$, and the matrix $\boldsymbol L$ is
\begin{eqnarray}
\begin{aligned}
\boldsymbol L&=\left[\begin{array}{cccccc}\label{eqA3}
L_{11} & L_{12} & L_{13} & L_{14} & L_{15} & L_{16} \\
L_{21} & L_{22} & L_{23} & L_{24} & L_{25} & L_{26} \\
L_{31} & L_{32} & L_{33} & L_{34} & L_{35} & L_{36} \\
L_{41} & L_{42} & L_{43} & L_{44} & L_{45} & L_{46} \\
L_{51} & L_{52} & L_{53} & L_{54} & L_{55} & L_{56} \\
L_{61} & L_{62} & L_{63} & L_{64} & L_{65} & L_{66}
\end{array}\right].
\end{aligned}
\end{eqnarray}
The matrix elements of $\boldsymbol L$ are
\begin{equation}
\begin{aligned}
L_{11}&=\frac{1}{2}\frac{\partial^2}{\partial x^2}+\mu+p-q-C_+\left(2n_{+1}+n_{0}\right)-C_-n_{-1}-\mathrm{i}v\frac{\partial}{\partial x},~~L_{12}=-C_+\varphi_{+1}^2, \\
L_{13}&=\frac{\mathrm{i}\gamma}{\sqrt{2}}\frac{\partial}{\partial x}-C_+\varphi_0^*\varphi_{+1}-2c_2\varphi_{-1}^*\varphi_{0},~~L_{14}=-C_+\varphi_0\varphi_{+1},~L_{15}=-C_-\varphi_{-1}^*\varphi_{+1},\\L_{16}&=-C_-\varphi_{-1}\varphi_{+1}-c_2\varphi_0^2,~~ L_{21}=C_+{\varphi_{+1}^*}^2,\\L_{22}&=-\frac{1}{2}\frac{\partial^2}{\partial x^2}-\mu-p+q+C_+\left(2n_{+1}+n_{0}\right)+C_-n_{-1}-\mathrm{i}v\frac{\partial}{\partial x},~~L_{23}=C_+\varphi_0^*\varphi_{+1}^*,\\L_{24}&=\frac{\mathrm{i}\gamma}{\sqrt{2}}\frac{\partial}{\partial x}+C_+\varphi_0\varphi_{+1}^*+2c_2\varphi_{-1}\varphi_{0}^*,~~L_{25}=C_-\varphi_{-1}^*\varphi_{+1}^*+c_2{\varphi_0^*}^2,~~L_{26}=C_-\varphi_{-1}\varphi_{+1}^*,\\
L_{31}&=\frac{\mathrm{i}\gamma}{\sqrt{2}}\frac{\partial}{\partial x}-C_+\varphi_0\varphi_{+1}^*-2c_2\varphi_{-1}\varphi_{0}^*,~~L_{32}=-C_+\varphi_0\varphi_{+1},\\L_{33}&=\frac{1}{2}\frac{\partial^2}{\partial x^2}+\mu-C_+\left(n_{+1}+n_{-1}\right)-2c_0n_{0}-\mathrm{i}v\frac{\partial}{\partial x},~~L_{34}=-2c_2\varphi_{-1}\varphi_{+1}-c_0\varphi_0^2,\\
L_{35}&=\frac{\mathrm{i}\gamma}{\sqrt{2}}\frac{\partial}{\partial x}-C_+\varphi_0\varphi_{-1}^*-2c_2\varphi_{+1}\varphi_{0}^*,~~L_{36}=-C_+\varphi_{-1}\varphi_0,~~
L_{41}=C_+\varphi_{+1}^*\varphi_0^*,\\L_{42}&=\frac{\mathrm{i}\gamma}{\sqrt{2}}\frac{\partial}{\partial x}+C_+\varphi_{+1}\varphi_{0}^*+2c_2\varphi_{0}\varphi_{-1}^*,~~L_{43}=2c_2\varphi_{-1}^*\varphi_{+1}^*+c_0{\varphi_0^*}^2,\\L_{44}&=-\frac{1}{2}\frac{\partial^2}{\partial x^2}-\mu+C_+\left(n_{+1}+n_{-1}\right)+2c_0n_{0}-\mathrm{i}v\frac{\partial}{\partial x},~~L_{45}=C_+\varphi_{-1}^*\varphi_0^*,\\
L_{46}&=\frac{\mathrm{i}\gamma}{\sqrt{2}}\frac{\partial}{\partial x}+C_+\varphi_{-1}\varphi_{0}^*+2c_2\varphi_{0}\varphi_{+1}^*,~~L_{51}=-C_-\varphi_{+1}^*\varphi_{-1},~~L_{52}=-C_-\varphi_{-1}\varphi_{+1}-c_2\varphi_0^2,\\
L_{53}&=\frac{\mathrm{i}\gamma}{\sqrt{2}}\frac{\partial}{\partial x}-C_+\varphi_0^*\varphi_{-1}-2c_2\varphi_{+1}^*\varphi_{0},~~L_{54}=-C_+\varphi_0\varphi_{-1},\\
L_{55}&=\frac{1}{2}\frac{\partial^2}{\partial x^2}+\mu-p-q-C_+\left(2n_{-1}+n_{0}\right)-C_-n_{+1}-\mathrm{i}v\frac{\partial}{\partial x},~~L_{56}=-C_+\varphi_{-1}^2,\\
L_{61}&=C_-\varphi_{-1}^*\varphi_{+1}^*+c_2{\varphi_0^*}^2,~~L_{62}=C_-\varphi_{+1}\varphi_{-1}^*,~~
L_{63}=C_+\varphi_{0}^*\varphi_{-1}^*,\\L_{64}&=\frac{\mathrm{i}\gamma}{\sqrt{2}}\frac{\partial}{\partial x}+C_+\varphi_0\varphi_{-1}^*+2c_2\varphi_{+1}\varphi_{0}^*,~~L_{65}=C_+{\varphi_{-1}^*}^2,\\L_{66}&=-\frac{1}{2}\frac{\partial^2}{\partial x^2}-\mu+p+q+C_+\left(2n_{-1}+n_{0}\right)+C_-n_{+1}-\mathrm{i}v\frac{\partial}{\partial x}.
\end{aligned}
\end{equation}
Here, $C_+=c_0+c_2$ and $C_-=c_0-c_2$. Then we transform the equation~\eqref{eqA3} into the momentum space, and solve the eigenvalues $\lambda$ numerically. We can judge the stability of solitons based on whether the real part maximum of $\lambda$ is greater than $10^{-3}$.


\section*{References}
\begin{thebibliography}{99} 
\bibitem{lin_spinorbit-coupled_2011}Lin Y-J, Jiménez-García K and Spielman I B 2011 Spin–orbit-coupled Bose–Einstein condensates {\it Nature} {\bf 471} 83–6

\bibitem{luo_tunable_2016}Luo X, Wu L, Chen J, Guan Q, Gao K, Xu Z-F, You L and Wang R 2016 Tunable atomic spin-orbit coupling synthesized with a modulating gradient magnetic field {\it Sci. Rep.} {\bf 6} 18983

\bibitem{bychkov_oscillatory_1984}Bychkov Y A and Rashba E I 1984 Oscillatory effects and the magnetic susceptibility of carriers in inversion layers {\it J. Phys. C: Solid State Phys.} {\bf 17} 6039–45

\bibitem{dresselhaus_spin-orbit_1955}Dresselhaus G 1955 Spin-Orbit Coupling Effects in Zinc Blende Structures {\it Phys. Rev.} {\bf 100} 580–6

\bibitem{qi_topological_2011}Qi X-L and Zhang S-C 2011 Topological insulators and superconductors {\it Rev. Mod. Phys.} {\bf 83} 1057–110

\bibitem{sinova_spin_2015}Sinova J, Valenzuela S O, Wunderlich J, Back C H and Jungwirth T 2015 Spin Hall effects {\it Rev. Mod. Phys.} {\bf 87} 1213–60

\bibitem{wang_spin-orbit_2010}Wang C, Gao C, Jian C-M and Zhai H 2010 Spin-Orbit Coupled Spinor Bose-Einstein Condensates {\it Phys. Rev. Lett.} {\bf 105} 160403

\bibitem{wen_ground_2012}Wen L, Sun Q, Wang H Q, Ji A C and Liu W M 2012 Ground state of spin-1 Bose-Einstein condensates with spin-orbit coupling in a Zeeman field {\it Phys. Rev.} A {\bf86} 043602

\bibitem{ho_bose-einstein_2011}Ho T-L and Zhang S 2011 Bose-Einstein Condensates with Spin-Orbit Interaction {\it Phys. Rev. Lett.} {\bf 107} 150403

\bibitem{li_quantum_2012}Li Y, Pitaevskii L P and Stringari S 2012 Quantum Tricriticality and Phase Transitions in Spin-Orbit Coupled Bose-Einstein Condensates {\it Phys. Rev. Lett.} {\bf 108} 225301

\bibitem{lan_raman-dressed_2014}Lan Z and Öhberg P 2014 Raman-dressed spin-1 spin-orbit-coupled quantum gas {\it Phys. Rev.} A {\bf 89} 023630

\bibitem{natu_striped_2015}Natu S S, Li X and Cole W S 2015 Striped ferronematic ground states in a spin-orbit-coupled $S = 1$ Bose gas {\it Phys. Rev.} A {\bf 91} 023608

\bibitem{sun_interacting_2016}Sun K, Qu C, Xu Y, Zhang Y and Zhang C 2016 Interacting spin-orbit-coupled spin-1 Bose-Einstein condensates {\it Phys. Rev.} A {\bf 93} 023615

\bibitem{campbell_magnetic_2016}Campbell D L, Price R M, Putra A, Valdés-Curiel A, Trypogeorgos D and Spielman I B 2016 Magnetic phases of spin-1 spin–orbit-coupled Bose gases {\it Nat. Commun.} {\bf 7} 10897

\bibitem{maeland_plane-_2020}Mæland K, Janssønn A T G, Rygh J H and Sudbø A 2020 Plane- and stripe-wave phases of a spin-orbit-coupled Bose-Einstein condensate in an optical lattice with a Zeeman field {\it Phys. Rev.} A {\bf 102} 053318

\bibitem{adhikari_multiring_2021}Adhikari S K 2021 Multiring, stripe, and superlattice solitons in a spin-orbit-coupled spin-1 condensate {\it Phys. Rev.} A {\bf 103} L011301

\bibitem{adhikari_symbiotic_2021}Adhikari S K 2021 Symbiotic solitons in quasi-one- and quasi-two-dimensional spin-1 condensates {\it Phys. Rev.} E {\bf 104} 024207

\bibitem{chen_elementary_2022}Chen Y, Lyu H, Xu Y and Zhang Y 2022 Elementary excitations in a spin–orbit-coupled spin-1 Bose–Einstein condensate {\it New J. Phys.} {\bf 24} 073041

\bibitem{li_stripe_2017}Li J-R, Lee J, Huang W, Burchesky S, Shteynas B, Top F Ç, Jamison A O and Ketterle W 2017 A stripe phase with supersolid properties in spin–orbit-coupled Bose–Einstein condensates {\it Nature} {\bf 543} 91–4

\bibitem{sanchez-baena_supersolid_2020}Sánchez-Baena J, Boronat J and Mazzanti F 2020 Supersolid striped droplets in a Raman spin-orbit-coupled system {\it Phys. Rev.} A {\bf 102} 053308

\bibitem{wan_solitons_2019}Wan N-S, Li Y-E and Xue J-K 2019 Solitons in spin-orbit-coupled spin-2 spinor Bose-Einstein condensates {\it Phys. Rev.} E {\bf 99} 062220

\bibitem{wen_motion_2016}Wen L, Sun Q, Chen Y, Wang D-S, Hu J, Chen H, Liu W-M, Juzeliūnas G, Malomed B A and Ji A-C 2016 Motion of solitons in one-dimensional spin-orbit-coupled Bose-Einstein condensates {\it Phys. Rev.} A {\bf 94} 061602

\bibitem{kartashov_bloch_2016}Kartashov Y V, Konotop V V, Zezyulin D A and Torner L 2016 Bloch Oscillations in Optical and Zeeman Lattices in the Presence of Spin-Orbit Coupling {\it Phys. Rev. Lett.} {\bf 117} 215301

\bibitem{kartashov_solitons_2019}Kartashov Y V, Konotop V V, Modugno M and Sherman E Ya 2019 Solitons in Inhomogeneous Gauge Potentials: Integrable and Nonintegrable Dynamics {\it Phys. Rev. Lett.} {\bf 122} 064101

\bibitem{qiu_dynamics_2023}Qiu X, Hu A-Y, Cai Y, Saito H, Zhang X-F and Wen L 2023 Dynamics of spin-nematic bright solitary waves in spin-tensor-momentum coupled Bose-Einstein condensates {\it Phys. Rev.} A {\bf 107} 033308

\bibitem{xu_bright_2015}Xu Y, Zhang Y and Zhang C 2015 Bright solitons in a two-dimensional spin-orbit-coupled dipolar Bose-Einstein condensate {\it Phys. Rev.} A {\bf 92} 013633

\bibitem{xu_bright_2013}Xu Y, Zhang Y and Wu B 2013 Bright solitons in spin-orbit-coupled Bose-Einstein condensates {\it Phys. Rev.} A {\bf 87} 013614

\bibitem{achilleos_matter-wave_2013}Achilleos V, Frantzeskakis D J, Kevrekidis P G and Pelinovsky D E 2013 Matter-Wave Bright Solitons in Spin-Orbit Coupled Bose-Einstein Condensates {\it Phys. Rev. Lett.} {\bf 110} 264101

\bibitem{kartashov_bose-einstein_2014}Kartashov Y V, Konotop V V and Zezyulin D A 2014 Bose-Einstein condensates with localized spin-orbit coupling: Soliton complexes and spinor dynamics {\it Phys. Rev.} A {\bf 90} 063621

\bibitem{sakaguchi_vortex_2016}Sakaguchi H, Sherman E Ya and Malomed B A 2016 Vortex solitons in two-dimensional spin-orbit coupled Bose-Einstein condensates: Effects of the Rashba-Dresselhaus coupling and Zeeman splitting {\it Phys. Rev.} E {\bf 94} 032202

\bibitem{li_moving_2016}Li Y-E and Xue J-K 2016 Moving Matter-Wave Solitons in Spin—Orbit Coupled Bose—Einstein Condensates {\it Chin. Phys. Lett.} {\bf 33} 100502

\bibitem{kartashov_multidimensional_2020}Kartashov Y V, Torner L, Modugno M, Sherman E Ya, Malomed B A and Konotop V V 2020 Multidimensional hybrid Bose-Einstein condensates stabilized by lower-dimensional spin-orbit coupling {\it Phys. Rev. Res.} {\bf 2} 013036

\bibitem{kartashov_solitons_2017}Kartashov Y V and Konotop V V 2017 Solitons in Bose-Einstein Condensates with Helicoidal Spin-Orbit Coupling {\it Phys. Rev. Lett.} {\bf 118} 190401

\bibitem{li_solitary_2021}Li X-X, Cheng R-J, Ma J-L, Zhang A-X and Xue J-K 2021 Solitary matter wave in spin-orbit-coupled Bose-Einstein condensates with helicoidal gauge potential {\it Phys. Rev.} E {\bf 104} 034214

\bibitem{zhao_topological_2015}Zhao D, Song S-W, Wen L, Li Z-D, Luo H-G and Liu W-M 2015 Topological defects and inhomogeneous spin patterns induced by the quadratic Zeeman effect in spin-1 Bose-Einstein condensates {\it Phys. Rev.} A {\bf 91} 013619

\bibitem{zapf_bose-einstein_2014}Zapf V, Jaime M and Batista C D 2014 Bose-Einstein condensation in quantum magnets {\it Rev. Mod. Phys.} {\bf 86} 563–614

\bibitem{sun_bright_2020}Sun J, Chen Y, Chen X and Zhang Y 2020 Bright solitons in a spin-tensor-momentum-coupled Bose-Einstein condensate {\it Phys. Rev.} A {\bf 101} 053621

\bibitem{tononi_quantum_2019}Tononi A, Wang Y and Salasnich L 2019 Quantum solitons in spin-orbit-coupled Bose-Bose mixtures {\it Phys. Rev.} A {\bf 99} 063618

\bibitem{kawaguchi_spinor_2012}Kawaguchi Y and Ueda M 2012 Spinor Bose–Einstein condensates {\it Phys. Rep.} {\bf 520} 253–381

\bibitem{gautam_mobile_2015}Gautam S and Adhikari S K 2015 Mobile vector soliton in a spin–orbit coupled spin-1 condensate {\it Laser Phys. Lett.} {\bf 12} 045501

\bibitem{adhikari_phase_2019}Adhikari S K 2019 Phase separation of vector solitons in spin-orbit-coupled spin-1 condensates {\it Phys. Rev.} A {\bf 100} 063618

\bibitem{gautam_vortex-bright_2017}Gautam S and Adhikari S K 2017 Vortex-bright solitons in a spin-orbit-coupled spin-1 condensate {\it Phys. Rev.} A {\bf 95} 013608

\bibitem{he_multi-type_2022}He J-T, Fang P-P and Lin J 2022 Multi-Type Solitons in Spin-Orbit Coupled Spin-1 Bose–Einstein Condensates {\it Chin. Phys. Lett.} {\bf 39} 020301

\bibitem{qi_soliton_2023}Qi J, Zhao D and Liu W-M 2023 Soliton collisions in spin–orbit coupled spin-1 Bose–Einstein condensates {\it J. Phys. A: Math. Theor.} {\bf 56} 255702

\bibitem{li_stationary_2018}Li Y-E and Xue J-K 2018 Stationary and moving solitons in spin–orbit-coupled spin-1 Bose–Einstein condensates {\it Front. Phys.} {\bf 13} 130307

\bibitem{papoular_microwave-induced_2010}Papoular D J, Shlyapnikov G V and Dalibard J 2010 Microwave-induced Fano-Feshbach resonances {\it Phys. Rev.} A {\bf 81} 041603

\bibitem{ravisankar_effect_2021}Ravisankar R, Fabrelli H, Gammal A, Muruganandam P and Mishra P K 2021 Effect of Rashba spin-orbit and Rabi couplings on the excitation spectrum of binary Bose-Einstein condensates {\it Phys. Rev.} A {\bf 104} 053315

\bibitem{kanna_exact_2003}Kanna T and Lakshmanan M 2003 Exact soliton solutions of coupled nonlinear Schrödinger equations: Shape-changing collisions, logic gates, and partially coherent solitons {\it Phys. Rev.} E {\bf 67} 046617

\bibitem{fu_nonlinear_2022}Fu Q, Wang P, Kartashov Y V, Konotop V V and Ye F 2022 Nonlinear Thouless Pumping: Solitons and Transport Breakdown {\it Phys. Rev. Lett.} {\bf 128} 154101

\bibitem{helm_sagnac_2015}Helm J L, Cornish S L and Gardiner S A 2015 Sagnac Interferometry Using Bright Matter-Wave Solitons {\it Phys. Rev. Lett.} {\bf 114} 134101

\bibitem{zhao_tunneling_2017}Zhao L-C, Ling L, Yang Z-Y and Yang W-L 2017 Tunneling dynamics between atomic bright solitons {\it Nonlinear Dyn.} {\bf 88} 2957–67

\bibitem{jia_soliton_2022}Jia Q, Qiu H and Mateo A M 2022 Soliton collisions in Bose-Einstein condensates with current-dependent interactions {\it Phys. Rev.} A {\bf 106} 063314
\end{thebibliography}
\end{document}